\documentclass[10pt, letter,twocolumn]{IEEEtran}

\usepackage{color}
\usepackage{url}
\usepackage{pdfsync}
\usepackage{epsf}
\usepackage{times}
\usepackage{epsfig}
\usepackage{graphicx}
\usepackage{algorithm}
\usepackage{algpseudocode}
\usepackage{algpseudocode}
\usepackage{amsmath}
\usepackage{amssymb}
\usepackage{amsxtra}
\usepackage{mathtools}
\usepackage{amsfonts}
\usepackage{cite}
\usepackage{cleveref}

\usepackage{bbm}
\usepackage{subfigure}
\usepackage{caption}
\usepackage{multirow}
\usepackage{array}
\usepackage{microtype}
\newcolumntype{P}[1]{>{\centering\arraybackslash}p{#1}}

\begin{document}
\title{A Hybrid Energy Sharing Framework for Green Cellular Networks\vspace{-0.0cm}}
\author{\IEEEauthorblockN{Muhammad Junaid Farooq, \textit{Student Member, IEEE}, \large Hakim Ghazzai, \textit{Member, IEEE}, Abdullah Kadri, \textit{Senior Member, IEEE}, Hesham ElSawy, \textit{Member, IEEE}, and Mohamed-Slim Alouini, \textit{Fellow, IEEE}\vspace{-0.0in}}\\

\thanks{\vspace{-0.0cm}\hrule \vspace{0.0cm}
\noindent A part of this work has been accepted for publication in IEEE Global Communications Conference (Globecom~2016).

\noindent This work was made possible by grant NPRP \# 6-001-2-001 from the Qatar National Research Fund (a member of The Qatar Foundation). The statements made herein are solely the responsibility of the authors.

\noindent Muhammad Junaid Farooq, Hakim Ghazzai, and Abdullah Kadri are with Qatar Mobility Innovations Center (QMIC), Qatar University, Doha, Qatar. E-mails: \{junaidf, hakimg, abdullahk\}@qmic.com.

\noindent Hesham ElSawy and Mohamed-Slim Alouini are with King Abdullah University of Science and Technology (KAUST), Thuwal, Makkah Province, Saudi Arabia. E-mails: \{hesham.elsawy, slim.alouini\}@kaust.edu.sa.

\noindent This work was done while Muhammad Junaid Farooq was at QMIC. He is now at New York University, NY, United States.
}

}

\maketitle
\thispagestyle{empty}
\begin{abstract}
\vspace{-0.0cm}
\boldmath
Cellular operators are increasingly turning towards renewable energy (RE) as an alternative to using traditional electricity in order to reduce operational expenditure and carbon footprint. Due to the randomness in both RE generation and mobile traffic at each base station (BS), a surplus or shortfall of energy may occur at any given time. To increase energy self-reliance and minimize the network's energy cost, the operator needs to efficiently exploit the RE generated across all BSs. In this paper, a hybrid energy sharing framework for cellular network is proposed, where a combination of physical power lines and energy trading with other BSs using smart grid is used. Algorithms for physical power lines deployment between BSs, based on average and complete statistics of the net RE available, are developed. Afterwards, an energy management framework is formulated to optimally determine the quantities of electricity and RE to be procured and exchanged among BSs, respectively, while considering battery capacities and real-time energy pricing. Three cases are investigated where RE generation is unknown, perfectly known, and partially known ahead of time. Results investigate the time varying energy management of BSs and demonstrate considerable reduction in average energy cost thanks to the hybrid energy sharing scheme.
\end{abstract}
\vspace{-0.0in}

\begin{IEEEkeywords}
\vspace{-0.0cm}
Cellular networks, clustering, energy sharing, physical power lines, smart grid.
\end{IEEEkeywords}

\vspace{-0.0cm}
\section{Introduction}
Cellular networks are growing at an ever increasing pace as the number of mobile phone users is rising rapidly due to ubiquitous connectivity needs. The demand is expected to continue increasing in the future as smart phones and other wireless devices become affordable for everyone~\cite{cisco_report}. To cater for this demand, more and more cellular base stations (BSs) are being installed in order to increase capacity, coverage, and quality of service (QoS). This is leading to a massive inflation in energy consumption of cellular networks, which is an alarming concern for cellular operators since energy costs constitute a significant portion of their operational expenditures (OPEX)~\cite{opex}. Additionally, the prices of information and communication technology (ICT) services worldwide are falling gradually~\cite{falling_prices}. This is further increasing the pressure on cellular operators to reduce energy consumption in order to maintain profitability. Apart from the financial implications, the rising energy consumption is also contributing negatively towards the environment due to its association with combustion of fossil fuels~\cite{footprint_of_communication}. Therefore, cellular operators and equipment manufacturers are both focused towards curbing the growing energy consumption of the BSs~\cite{energy_consumption2} and increasing self-reliance of energy using renewable sources~\cite{flatten_energy}.

Most cellular operators worldwide are deploying renewable energy (RE) generators, e.g., solar panels, wind turbines, etc., at the sites of their BSs to obtain clean and cheap electricity~\cite{solar_powered_BS,solar_installation,suggested3}. These RE generators are used in conjunction with traditional electricity and backup energy storage units to power up the BSs. The amount of RE generated is highly variable in time and space depending on different factors. For instance, solar energy depends on the amount of sunlight, panel area, and energy conversion efficiency, etc. Hence, the BSs generating energy from renewable sources may sometimes have a surplus energy while at other times, they may be energy deficient. The energy deficit may be complemented by purchasing electricity from smart grid (SG) while any surplus RE may be sold back, thanks to the intelligent two-way power flow enabled by the SG~\cite{smart_grid,7570259}. However, it will be more cost effective if the distributed RE generated at BSs sites is synergized by a common energy infrastructure to collectively serve the energy requirements of all BSs~\cite{micro_grid,micro-grid2,nirwan_ansari}. In other words, the excess RE at one BS will compensate for the deficit at another BS by enabling energy sharing between the two BSs.

Energy sharing allows higher utilization of locally generated RE by the BSs and helps in further curtailing the energy cost by reducing procurement from SG. However, the realization of energy sharing among BSs requires an effective energy transport mechanism. SG may be used for virtually transporting energy between BSs by selling extra energy at one BS and buying the same amount of energy at another BS at a preferential price. However, SG imposes a charge for providing this service that is reflected by the difference between the buying and selling prices. Secondly, the energy cost, and hence the operator's revenues, are sensitive to variation in SG pricing policies. Another approach for energy sharing could be the connection of BSs by physical power lines. Although this is infeasible for long distances due to high installation costs, resistive power losses, and right-of-way restriction, etc., it may be plausible to install physical power lines for sharing energy among BSs within small localities. In this manner, the operator will not incur any additional cost for sharing energy within short distances and its profits will be more robust to the variation of SG energy trading prices. For long distances where physical cabling might not be possible, SG infrastructure may be used for energy transfer among BSs.

\textcolor{black}{The physical power lines essentially eliminate the cost of energy sharing among BSs via the SG at the cost of high initial capital investment. However, despite the initial investment required for deploying physical power lines, financial gains from energy sharing might outweigh costs in the long-term. Nevertheless, it is still important to precisely determine which BSs should be connected by physical power lines and which BSs should virtually share energy via SG in order to minimize the capital expenditure. Note that the connections among BSs using physical power lines are permanent. Hence, the installations must be carefully planned after making a thorough assessment of energy requirements.}

\vspace{-0.0cm}
\subsection{Related Work}
\textcolor{black}{Several studies focusing on reducing the conventional electricity usage of cellular networks exist in literature. In fact, different techniques have been used to achieve energy efficiency such as infrastructure sharing~\cite{suggested2,infrastructure_sharing_recent}, BS ON/OFF switching~\cite{forced_reference_1,on_off_recent,switching_off_recent}, radio resource management~\cite{resource_management_recent,hakim_vtc}, and energy harvesting from renewable sources to power BSs~\cite{solar_powered_BS,solar_installation,suggested3}.
In~\cite{suggested2,infrastructure_sharing_recent}, infrastructure sharing approaches are proposed where multiple cellular operators use the same radio access network infrastructure to reduce the otherwise redundant capacity of BSs that results in energy inefficiency. On the other hand,\cite{forced_reference_1,on_off_recent,switching_off_recent} propose different strategies based on turning off underutilized BSs and increasing the coverage of the remaining BSs to improve energy utilization of the network.
In~\cite{resource_management_recent,hakim_vtc}, energy efficient radio resource management frameworks are proposed for heterogeneous networks that employ multiple radio access technologies.
However, the most promising and currently deployed energy-efficient solutions are based on the use of RE to power cellular networks~\cite{solar_powered_BS,solar_installation,suggested3}. The benefit of using RE in cellular networks has been thoroughly investigated in literature ~\cite{bhargava,energy_management1,procurement,smart_grid_cellular,sharing1}. RE has shown to yield significant OPEX and carbon dioxide (CO$_2$) reductions for cellular operators by reducing reliance on traditional electricity supply~\cite{bhargava}.} In addition to employing RE sources, efficient energy management techniques are required to optimize the energy cost of the operator via SG~\cite{energy_management1,procurement,drm_junaid,smart_grid_cellular}. In~\cite{energy_management1}, an optimal energy management strategy is proposed to reduce the energy cost of a single SG connected BS with RE generation and battery storage. \cite{procurement} and \cite{drm_junaid} optimize the energy procurement of the operator and demand response from the suppliers respectively in the presence of multiple electricity retailers in SG to reduce cost as well as environmental impact. However, they do not account for the time varying effect of RE generation, user traffic, and electricity prices. On the other hand, in~\cite{smart_grid_cellular}, the OPEX reduction is achieved by a combination of turning off BSs dynamically and efficient energy procurement from retailers in SG based on the traffic and real-time electricity prices. These frameworks attempt to achieve cost efficiency for each BS individually without any energy interactions among BSs.

Recently, the concept of energy sharing among BSs has emerged as another step towards self-reliance of cellular networks in terms of energy~\cite{energy_exchange1,globecom,sharing3}. This paradigm has also been referred to as \emph{energy exchange}~\cite{energy_exchange1} and \emph{energy cooperation}~\cite{sharing3,sharing1} in the literature. Energy sharing is motivated by the fact that the distributed RE generated at BSs sites can be connected together to create a microgrid~\cite{microgrid_def} that collectively serves the energy requirements of all BSs. A novel architecture for microgrid connected green cellular networks is proposed in~\cite{micro_grid} where the REs generated at BSs sites and at centralized locations are integrated to jointly serve all BSs. 
A framework for energy exchange among BSs using SG is proposed in~\cite{energy_exchange1}. \textcolor{black}{However, it presents an off-line strategy and does not consider the uncertainty in RE generation. Similarly, in~\cite{globecom}, an optimized energy management framework is proposed for microgrid-connected cellular BSs that are equipped with RE generators and battery storage. It uses the SG as the common energy infrastructure and considers time variations and uncertainty in RE generation in the optimization. However, there is no energy sharing using physical power lines.} On the other hand, an optimized energy management framework for physically connected BSs that share energy is proposed in~\cite{sharing3}. However, it studies the case of two BSs that are connected by resistive power lines and generalizes the model for the larger network. This may not be feasible in practice since connecting each BS with other BSs in a mesh configuration using physical power lines is economically unviable and hence, not well-suited for large-scale cellular networks.

\vspace{-0.0cm}
\subsection{Contributions}
In this paper, a generalized energy sharing framework is developed for green cellular networks that use a combination of physical power lines among BSs and SG infrastructure to share locally generated RE. Particularly, we emphasize the need to intelligently install physical connections among BSs before overlaying them with virtual connections via SG. To the best of our knowledge, this hybrid framework for energy sharing in cellular networks has not been previously proposed or investigated. The main contributions of this paper are summarized as follows:
\begin{enumerate}
  \item A generalized hybrid energy sharing framework is developed for cellular BSs possessing RE generation and storage capabilities.
  \item Pre-planned deployment of physical power lines among BSs for sharing energy is proposed based on average and complete statistics of RE generation and energy consumption of BSs. Agglomerative and divisive hierarchical clustering algorithms are developed to determine the physical energy sharing links among the BSs. Two metrics are proposed for each algorithm namely the average energy affinity (AEA) and stochastic energy affinity (SEA). The AEA metric is based on the average energy comparisons while the SEA metric is based on a probabilistic comparison.
  \item An optimized energy management framework is proposed for sharing energy among BSs via SG while taking into account the already established physical connections. The energy management solutions are based on the level of \emph{a priori} knowledge of the system about RE generation, e.g., zero, perfect, and partial knowledge.
\end{enumerate}

The remainder of this paper is organized as follows: Section~\ref{sec_sys_model} presents the system model and provides an outline of the proposed energy sharing methodology. The algorithms to obtain physical connections among BSs are described in Section~\ref{sec_clustering} while the framework for optimized day-to-day operation is provided in Section~\ref{sec_day_to_day}. Numerical results are presented and explained in Section~\ref{sec_results}. Finally, the paper is concluded in Section~\ref{sec_conclusion}.
\vspace{-0.0cm}
\section{System Model \& Methodology} \label{sec_sys_model}
In this section, we present the proposed system model followed by an elucidation of the energy sharing concept.
\vspace{-0.0cm}
\subsection{System Model}
\begin{figure}
  \centering
  \includegraphics[width=3.4in]{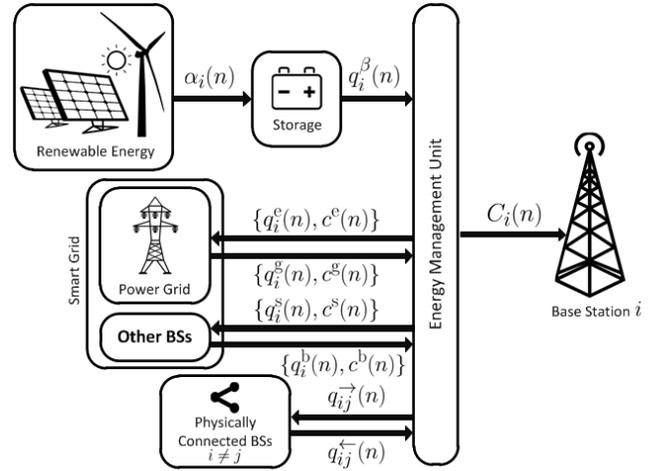}\vspace{-0.0in}
  \caption{System model for individual BSs.}\label{sys_model} \vspace{-0.0in}
\end{figure}

We consider a cellular network comprising of $K$ BSs located in an $L$ $\times$ $L$ km$^2$ square region in $\mathbb{R}^2$. \textcolor{black}{The coordinate vectors representing the locations of the BSs are denoted by $\mathbf{X}_{i}, \forall i \in \{1,\ldots,K\}$}. We assume a time slotted system with $N$ slots where $n \in \{1,\ldots,N\}$ represents the index of the time slot. Fig.~\ref{sys_model} shows the energy sharing model for a typical BS in the network. A typical BS possesses RE generation potential in the form of solar panels and/or wind turbines. The RE generated at BS $i \in \{1,\ldots,K\}$ during time slot $n$ is a random variable denoted by $\alpha_{i}(n) \in [0,\alpha^{\max}_{i}]$, where $\alpha^{\max}_{i}$ is the maximum generation capacity available. On the other hand, energy consumption of BSs is also a random variable denoted by $C_{i}(n) \in [0,C_{i}^{\max}], \forall i \in \{1,\ldots,K\}$. The net renewable energy available (NRE) at a typical BS, denoted by $E_{i}(n)$, is the difference between $\alpha_{i}(n)$ and $C_{i}(n)$. A positive $E_{i}$ indicates that BS $i$ has surplus RE while negative $E_{i}(n)$ indicates that it has an energy deficit. Due to the randomness in energy generation and energy consumption of each BS, the NRE at a typical BS is also a random variable over the support $[-C_{i}^{\max}, \alpha_{i}^{\max}]$ with a probability density function $f_{E_{i}}(E_{i})$. \textcolor{black}{Note that the energy consumption of BSs might be correlated due to the spatial distribution of users or the use of cooperative communication techniques~\cite{reviewer4_suggested}. Moreover, the RE generation is also spatially correlated due to similar environmental conditions observed by neighbouring BSs. Hence, the resulting NREs of the BSs are not independent in general.} The energy quantities used in this paper, \textcolor{black}{e.g., $\alpha_{i}(n)$, are obtained from their corresponding continuous time power profiles, e.g., $\alpha_{i}^{\prime}(t)$ as follows: $\alpha_{i}(n) =  \int_{(n-1)\tau}^{n\tau} \alpha_{i}^{\prime}(t) dt$, where $\tau$ is the duration of each time slot. Hence, they represent the temporal averages of the energies over the corresponding time slots.}

Each BS is also connected to SG which provides access to both traditional electricity and other cellular BSs. SG is a flexible electricity grid which allows the integration of multiple energy sources and provides advanced energy management features~\cite{smart_grid}. In the considered system model, SG performs the following two main functions: (i) two-way power flow with the cellular BSs, i.e., the BS can procure electricity to meet its requirements and sell electricity in case it has surplus and unwanted RE, and (ii) virtual transport and aggregation of excess RE from a group of BSs to an energy deficient BS, i.e., BSs with surplus energies at different geographical locations can share the energy with an energy deficient BS at a preferential price.

The amount of traditional electricity obtained from SG by BS $i$ is denoted by $q_{i}^{\text{g}}(n)$ at a price of $c^{\text{g}}(n)$ monetary units (MU). The amount of extra electricity sold back to SG at a price of $c^{\text{e}}(n)$ MU is denoted by $q_{i}^{\text{e}}(n)$. \textcolor{black}{Note that the events of a BS procuring electricity from SG and of selling back excess electricity are mutually exclusive. This implies that the product of the energy procured by SG and the energy sold back to SG is zero for the same time instant, i.e., $q_{i}^{\text{g}}(n) q_{i}^{\text{e}}(n) = 0, \forall n \in \{1,\ldots,N\}$. Hence, any BS will either procure electricity from SG to meet its energy deficiency or sell any extra available energy back to SG at any particular time.} The amount of energy bought by BS $i$ from other BSs via SG during time slot $n$ is denoted by $q_{i}^{\text{b}}(n) =  \sum_{j=1, j \neq i}^{K} q_{j}^{\text{s}}(n)$ at a price of $c^{\text{b}}(n)$ MU, while the energy sold to other BSs via SG is denoted by $q_{i}^{\text{s}}(n) =  \sum_{j=1, j \neq i}^{K} q_{j}^{\text{b}}(n)$ at a price of $c^{\text{s}}(n)$ MU. Note that, generally, the order of energy prices is $c^{\text{g}}(n) \gg  c^{\text{b}}(n) \geq c^{\text{s}}(n) \geq c^{\text{e}}(n)$. This means that the traditional electricity is an expensive source of energy typically priced much higher than the price of energy obtained from other BSs. For the energy that needs to be shared with other BSs of the same cellular network, SG will buy it from the BSs at a lower price than the price at which it will sell at another location, i.e., $c^{\text{b}}(n) \geq c^{\text{s}}(n)$. The least price is paid for the extra energy sold by the BSs back to SG. \textcolor{black}{It is pertinent to mention here that if the energy buying and selling prices of the SG are similar, there is no incentive for the operator to install dedicated power lines for energy sharing. However, since the SG is operated by private entities beyond the control of the cellular operator, it is highly unlikely that the energy transport is offered free of charge. Therefore, the presence of physical power lines alongside the already available SG infrastructure may be beneficial in further reducing the energy costs of the cellular operator.}

Finally, each BS may also have direct connections with other BSs via physical power lines. This provides the flexibility of sharing energy with other BSs without any charges. However, there may be power losses which need to be taken into account. The amount of energy obtained by BS $i$ from BS $j$ using a physical power line is denoted by $q^{\leftarrow}_{ij}$ while the amount of energy supplied to BS $j$ by BS $i$ is denoted by $q^{\rightarrow}_{ij}$. Each BS also owns an energy storage unit, e.g., a battery, with maximum capacity of $B_{\max}$ for buffering the generated RE. The energy supplied by the respective batteries to the BSs during time slot $n$ are denoted by $q_{i}^{\beta}(n), \forall i \in \{1,\ldots, K\}$. The instantaneous charge level of the battery at BS $i$, denoted by $B_{i}(n)$, is expressed as follows:
\begin{equation}\label{batt_level}
\small
  B_{i}(n) = B_{i}(n-1) + \alpha_{i}(n) - q_{i}^{\beta}(n) - q_{i}^{\text{s}}(n) - \sum \limits_{j = 1}^{K} A_{ij}q^{\rightarrow}_{ij}(n)- q_{i}^{\text{e}}(n),
\end{equation}
where $B_{i}(n - 1)$ denotes the charge level of the battery in the previous time slot and the factor $A_{ij}$ is $1$ if BS $i$ and $j$ are connected by physical power lines and $0$ otherwise. The initialization condition for the batteries is $B_{i}(0) = B_{0}, \forall i \in \{1, \ldots,K\}$. Note that the battery stores energy obtained from renewable sources only and not from energy shared by other BSs. \textcolor{black}{Note that we are using a simplified linear model for the charging and discharging of the battery for ease of analysis. However, in practice, the batteries display non-linear charging and discharging behaviours with respect to the load}.
\vspace{-0.0cm}
\subsection{Energy Sharing among Base Stations}
The energy sharing between a pair of BSs can be implemented using the following methods:
\vspace{-0.0cm}
\subsubsection{\textbf{Smart Grid Interface}}
Since all BSs are connected to SG, it becomes an attractive candidate to be used for energy sharing. The two-way power flow and metering allows for energy to be sold at one geographical location and purchasing the equivalent energy at another location. We refer to this as a \emph{virtual connection} since any BS is able to share energy with another BS without having a direct link between them, however, this does not come for free. SG has a different pricing policy for buying and selling energy. The difference between these prices, i.e., $c^{\text{b}}(n) - c^{\text{s}}(n)$, accounts for any energy losses and the costs for aggregating and virtually transporting energy, incurred by SG. Although there is no upfront investment to enable energy sharing using this approach, it may be costly in the long-term, depending on the amount of energy shared.

\subsubsection{\textbf{Physical Power Lines}}
Another approach for sharing energy among BSs is to install physical power lines to connect them together, also referred to as \emph{physical connection}. The initial installation may be costly because of the price of conductor, digging costs, and right-of-way payments, etc. Moreover, resistive losses can be significant particularly for longer distances. The amount of energy lost in the conductor in the form of heat during sharing is a function of the amount of energy transferred and the length of the power lines and is evaluated as follows~\cite{joule_heating}:
\begin{equation}\label{P_loss}
\small
  E_{loss}(E,l) = \left( I^{2}R(l) \right) \tau = \frac{P^2 R(l)}{V^2} \tau = \frac{E^{2} R(l)}{V^{2} \tau},
\end{equation}
where $I$ is the current passing through the conductor, $R(l)$ is the resistance of the conductor of length $l$ km, $E$ is the amount of energy transferred, and $V$ is the root-mean-square (rms) voltage\footnote{Note that the transmission of electrical energy may be done at higher voltages
to reduce power losses. Therefore, additional voltage up-conversion and down-conversion equipment may be required at BSs sites to transmit at high voltages.}. The resistance is a function of the distance and can be expressed as $R(l) = \rho l$~\cite{joule_heating}, where $\rho$ is the specific resistance of the conductor in $\Omega$/m. This approach can only be employed in a relatively small geographical region to avoid high power losses. In the considered model, we assume that physical power lines can only be installed if two BSs are within a distance of $r$ km from each other. This distance is henceforth referred to as \emph{energy sharing range}. From an economic perspective, physical power lines can yield great benefits if they are planned rationally. This is because there is no external cost incurred by the operator while sharing energy among BSs and the operator is unaffected by SG pricing variations.


\begin{figure}
  \centering
  \includegraphics[width=3.0in]{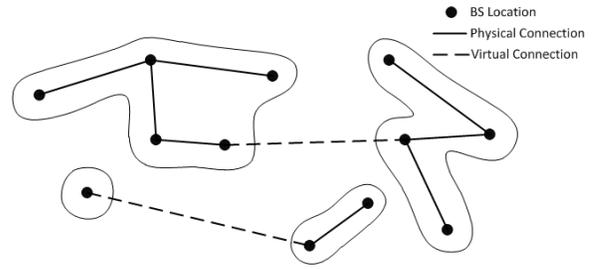}\vspace{-0.0in}
  \caption{Example of base stations connected by physical and virtual energy sharing links.}\label{ex1} \vspace{-0.1in}
\end{figure}
\vspace{-0.0cm}
\subsection{Methodology}

The objective of the proposed framework is to minimize the energy procurement cost of the cellular network by efficiently sharing RE among BSs via a combination of physical power lines and virtual connections using SG. An example of co-existing physical and virtual energy sharing links in a cellular network is shown in Fig.~\ref{ex1}. Virtual connections are flexible and can be added or removed frequently over time as the requirements change. However, physical connections are permanent and hence, need to be carefully planned before installation. Therefore, we first plan the physical connections between any two BSs based on long-term statistics and then develop a generalized framework that decides the energy procurement for day-to-day operations. We refer to the process of connecting the BSs using physical links as \emph{clustering}, where a cluster is defined by the set of nodes which are either connected directly or through other BSs in the graph as shown in Fig.~\ref{ex1}. We employ greedy graph clustering algorithms to determine the clusters with physical links. Once the clustering procedure is complete, the physical connections among BSs are assumed to be installed. Then, we develop an optimization problem for day-to-day cost minimization for the general case of having partial knowledge about RE generation. However, to provide a bench-mark, the cases of having zero knowledge and perfect knowledge about RE generation are also investigated.

%
\vspace{-0.0cm}
\section{Clustering Of Physically Connected Base Stations} \label{sec_clustering}

The first step in the development of an optimized energy sharing framework is to determine the set of links that can be realized by using physical power lines. Given the locations of the BSs and the statistics of the NRE available at each node, we can abstract the network into a graph with vertices representing the BSs and links representing the power lines. \textcolor{black}{Connecting the BSs together to form clusters from the graph as shown in Fig.~\ref{ex1}, without knowledge of the number of clusters, is an unsupervised machine learning problem and is non-deterministic polynomial-time hard (NP-hard)}. Hence, approximate heuristic algorithms are commonly used to achieve locally optimal results~\cite{algorithms}. We propose two algorithms inspired by the hierarchical clustering approaches to achieve an effective grouping of BSs that can significantly share energy most of the time. In the first algorithm, we begin from the case when all BSs are assumed to be disconnected from each other. The links are then iteratively added between two BSs to obtain a clustered graph. We refer to this as the \emph{Agglomerative Clustering}. The second algorithm begins with a mesh graph, where all BSs are assumed to be connected to each other. Redundant and unviable links are then successively removed to precisely obtain the final set of links that need to be installed. This is referred to as \emph{Divisive Clustering}. In the following subsections, we present the metrics that are used to assess the energy affinity of the BSs based on average and complete statistics as well as provide the algorithms to obtain the final set of energy sharing links.

\vspace{-0.0cm}
\subsection{Metrics for Linking BSs}

In order to decide which BSs will share energy using physical power lines, there is a need for a metric that accurately captures the energy affinity between one BS and other BSs. Since the NREs available at BSs are random over time and the physical links are installed on a long-term basis, it may not be reasonable to use instantaneous NRE statistics. Instead, we propose to use two metrics that are based on the amount of information available about the randomness of NRE. The two metrics are explained as follows:

\textbf{1- Average Energy Affinity (AEA):} In the event that only the average NRE available at each BS is known, we define AEA metric that measures the association between BS $i$ and BS $j$ based on the energy disparity and distance between these two BSs. The AEA metric is expressed by \eqref{wt_avg}.
\begin{figure*}
\begin{align}\label{wt_avg}
\small
\mathcal{M}^{\text{AEA}}_{ij} \hspace{-0.1cm} = \hspace{-0.1cm}
\left\{
	\begin{array}{ll}
        \hspace{-0.15cm}\max(\hat{E}_{i},\hat{E}_{j}) - \min(\hat{E}_{i},\hat{E}_{j}) - E_{\text{loss}}(\hat{\Pi}_{ij},\|\mathbf{X}_{i} - \mathbf{X}_{j} \|_{2}), \mbox{ \text{if} } \|\mathbf{X}_{i} - \mathbf{X}_{j}\|_{2} \leq r \text{ and } \text{sgn}(\hat{E}_{i}) \neq \text{sgn}(\hat{E}_{j}), \\
      \hspace{-0.15cm}  - \epsilon \|\mathbf{X}_{i} - \mathbf{X}_{j}\|_{2}, \hspace{1cm } \mbox{ \text{otherwise}},
\end{array}
\right.
\end{align}
\end{figure*}
where $\hat{E}_{i} = (1/N)\sum_{n = 1}^{N} \mathbb{E}[E_{i}(n)], \forall i \in \{ 1, \ldots, K\}$, represents the average NRE available at BS $i$ over all time slots and $\mathbb{E[.]}$ is the expectation operator. $\bar{\Pi}_{ij}$ represents the average energy transfer that takes place between BSs $i$ and $j$, if they are connected, and is expressed as $\bar{\Pi}_{ij} =		\min(|\hat{E}_{i}|,|\hat{E}_{j}|),$ where $|.|$ represents the absolute value and $\text{sgn}(.)$ represents the signum function that extracts the sign of a real number. Finally, the operator $\|.\|_{2}$ represents the Euclidean norm and hence, $\|\mathbf{X}_{i} - \mathbf{X}_{j}\|_{2}$ denotes the Euclidean distance between BSs $i$ and $j$. The AEA metric takes into account the difference between the average NRE available at the BSs and the amount of energy lost in the form of heat if the energy transfer takes place between them. Note that this metric takes different values depending upon the sign of the average NRE. If both BSs have either positive or negative average NREs, i.e., $\text{sgn}(\hat{E}_{i}) = \text{sgn}(\hat{E}_{j})$, then installing physical power line between them is not useful on average. Hence, to discourage the selection of such links, we assign an arbitrarily large negative constant weight, i.e., $-\epsilon$, and scale it by the distance between the BSs so that the worst links can still be ranked according to the distance. However, if signs of the average NRE of the two BSs are different, then the metric prioritizes the links based on the energy difference and the energy loss in the form of heat. A higher value of the metric indicates a high affinity between the two BSs to share energy and vice versa.

\textbf{2- Stochastic Energy Affinity (SEA):} In the event that complete statistics of the NREs of BSs are available, a stochastic metric measuring energy affinities between each pair of BSs may be more accurate. The SEA metric, based on probabilistic comparison of the NREs, is given by \eqref{wt_stochastic}.
\begin{figure*}
      \begin{equation}\label{wt_stochastic}
\small
      \mathcal{M}^{\text{SEA}}_{ij} =
      \left\{
    	\begin{array}{ll}
            \frac{1}{N} \sum \limits_{n=1}^{N} \mathbb{P}[|E_{i}(n) - E_{j}(n)| > \delta], &  \mbox{if } \|\mathbf{X}_{i} - \mathbf{X}_{j}\|_{2} \leq r \text{ and }  \left(\prod \limits_{n=1}^{N }\mathbb{P}[ \underset{k \in \{i,j\}}{\cap} \{ {E}_{k}(n) > 0 \} \ \ \text{or} \underset{k \in \{i,j\}}{\cap} \{ {E}_{k}(n) < 0 \}] \right)^{\frac{1}{N}} < \phi_{l}, \\
            \zeta \|\mathbf{X}_{i} - \mathbf{X}_{j}\|_{2},  & \mbox{otherwise}.
      \end{array}
      \right.
      \end{equation}
\end{figure*}
Instead of considering the difference in the average NRE between the BSs, the SEA metric considers the temporal average of the probability that NREs of the two BSs differ by more than a constant $\delta$. The constant $\delta$ controls the number of energy sharing links that will be installed as a higher value of $\delta$ will lead to fewer pairs of BSs qualifying for sharing energy and vice versa. The condition of the metric also becomes probabilistic and can be expressed as the geometric mean of the probability of NRE of both BSs being either positive or negative is less than a constant $\phi_{l}$. In fact, the constant $\phi_{l}$ controls the usefulness of the energy sharing link between the BSs. If the link is considered to be less useful, the metric assigns a weight of $\zeta$ scaled by the distance between the BSs, where $\zeta$ is an arbitrarily small constant approaching zero introduced to reflect the low importance of the link. Note that the geometric mean is used in the condition instead of the arithmetic mean to reduce the impact of extreme values on the metric. This is because a few time slots in which the probability of two BSs having NREs with opposite signs may completely prevent them to share energy if the arithmetic mean is used. However, the geometric mean is more robust to such rare fluctuations in the probabilities.

\vspace{-0.0cm}
\subsection{Clustering Algorithms}
The clustering algorithms developed in order to provide the best combination of physical power lines to connect the BSs are provided in this section. Since the number of clusters and the number of BSs are unknown \emph{a priori}, we use the unsupervised machine learning approach of hierarchical clustering~\cite{algorithms} that iteratively leads to the final set of clusters. Table~\ref{table_expressions} lists the expressions that are used in the algorithms for the case of AEA and SEA metrics and abbreviates them with labels. These labels are then used in the steps of the algorithms to avoid repetition. The agglomerative and divisive clustering algorithms are presented as follows:
\subsubsection{Agglomerative Approach}

\begin{table}[]
\centering

\caption{Abbreviations for expressions related to average and stochastic energy affinity metrics}\vspace{-0.1in}
\label{table_expressions}
\begin{tabular}{|c|c|c|}
\hline
\multirow{2}{*}{Label} & \multicolumn{2}{c|}{Expression} \\ \cline{2-3}
                       & AEA       & SEA      \\ \hline
\textbf{S1($i$)}     &       $\hat{E}_{i} < 0$        &      $\left(\prod_{n=1}^{N} \mathbb{P}[E_{i}(n) < 0]\right)^{\frac{1}{N}} > \phi_{h}$        \\ \hline
\textbf{S2($i$)}     &   $\underset{i \in \mathcal{E}^{-}}{\arg \min} \ \hat{E}_{i}$   &   $\underset{i \in \mathcal{E}^{-}}{\arg \max} \  \left(\prod_{n=1}^{N} \mathbb{P}[E_{i}(n) < 0]\right)^{\frac{1}{N}}$              \\ \hline
\textbf{S3($j$)}   &         $\underset{j \in \mathcal{N}_{\hat{i}}}{\max} \ \hat{E}_{j} < 0$      &       $\underset{j \in \mathcal{N}_{\hat{i}}}{\max} \left(  \left(\prod_{n=1}^{N} \mathbb{P}[E_{j}(n) < 0] \right)^{\frac{1}{N}} \right) > \phi_{h}$          \\ \hline
\textbf{S4($i,j$)}     &         $\mathcal{M}_{ij}^{\text{AEA}}$      &       $\mathcal{M}_{ij}^{\text{SEA}}$          \\ \hline
\end{tabular}\vspace{-0.0cm}
\end{table}

\begin{algorithm}[h]
\small
\renewcommand{\thealgorithm}{1a}
\caption{Agglomerative Clustering}
\label{alg1:Agglomerative_algo}
\begin{algorithmic}[1]
\Require BS locations $\mathbf{X}_{i}, \forall i \in \{1,\ldots,K\}$, NRE $E_{i}(n), \forall i \in \{1,\ldots,K\},n \in \{1,\ldots,N\}$.
\State Initialize the set of BSs that have negative NRE according to the condition \textbf{S1}$(i)$, $\mathcal{E}^{-}=\{i \in \{1,\ldots,K\} : \textbf{S1}(i)\}$ and the BS association matrix $\mathbf{A}= [\boldsymbol{0}]_{K \times K}$.
\Repeat \label{repeat_start}
\State Identify the BS with maximum energy deficit from the set $\mathcal{E}^{-}$, i.e., $\hat{i} =$ \textbf{S2}$(i)$.
\State Find BSs that are within a distance of $r$ km of BS $\hat{i}$, i.e., $\mathcal{N}_{\hat{i}} = \{ j \in \{1,\ldots,K\}\backslash \{\hat{i}\} : \|\mathbf{X}_{\hat{i}} - \mathbf{X}_{j} \|_{2} \leq r  \}$. \label{norm_notation}
\If { $\textbf{S3}(j)$ = 1, $\forall j \in \mathcal{N}_{\hat{i}}$,} \label{check_all_deficient}
\State BS $\hat{i}$ cannot be connected to any of its neighbours. Remove $\hat{i}$ from the set $\mathcal{E}^{-}$, i.e., $\{\mathcal{E}^{-}\} \gets \{\mathcal{E}^{-}\}\backslash \{\hat{i}\}$.\label{remove_BS_1}
\Else
\State Calculate the metric $\textbf{S4}(\hat{i},j)$, $\{ \forall j \in \mathcal{N}_{\hat{i}}: \textbf{S1}(j) \neq 1\}$.
\State Select the index of BS with the best metric $\hat{j} =  \underset{j \in \mathcal{N}_{\hat{i}}}{\arg \max} \ \textbf{S4}(\hat{i},j)$.
\State Connect BS $\hat{i}$ and $\hat{j}$ by updating the association matrix, i.e., $\mathbf{A}_{\hat{i}\hat{j}} = \mathbf{A}_{\hat{j}\hat{i}} \gets  1$.
\State Execute the energy update procedure described in \textbf{Algorithm}~\ref{alg1:update} for both AEA and SEA metrics.
\EndIf
\Until the set of energy deficient BSs $\mathcal{E}^{-}$ is empty. \label{repeat_end}
\State The final association matrix $\mathbf{A}$ represents the clusters of BSs that share energy by physical power lines.
\end{algorithmic}
\end{algorithm}
\normalsize

\begin{algorithm}
\small
\renewcommand{\thealgorithm}{1b}
\caption{Update in Algorithm 1}
\label{alg1:update}
\begin{algorithmic}[1]
\Procedure{AEA Update}{}\Comment{For AEA metric}
\If {$\mathcal{M}_{\hat{i}\hat{j}} > 0$}
\State Update net energy of energy donating BS, i.e., $\hat{E}_{\hat{j}} \gets \mathcal{M}_{\hat{i}\hat{j}}$.
\State Remove BS $\hat{i}$ from set of energy deficient BSs, i.e., $\{\mathcal{E}^{-}\} \gets \{\mathcal{E}^{-}\}\backslash \{\hat{i}\}$.
\Else
\State Update net energy of energy donating BS, i.e., $\hat{E}_{\hat{j}} \gets 0$.
\State Update net energy of energy deficient BS, i.e, $\hat{E}_{i} \gets \mathcal{M}_{\hat{i}\hat{j}}$.
\EndIf
\EndProcedure
\Procedure{SEA Update}{}\Comment{For SEA metric}
\State $E_{\hat{i}}(n) \gets E_{\hat{i}}(n) + \min(|\mathbb{E}[E_{\hat{i}}(n)]|,|\mathbb{E}[E_{\hat{j}}(n)]|), \forall n \in \{1,\ldots,N\}$.
\State $E_{\hat{j}}(n) \gets E_{\hat{j}}(n) - \min(|\mathbb{E}[E_{\hat{i}}(n)]|,|\mathbb{E}[E_{\hat{j}}(n)]|), \forall n \in \{1,\ldots,N\}$.
\EndProcedure
\end{algorithmic}
\end{algorithm}
\normalsize

In this approach, we begin with each BS as an independent cluster and then successively connect a pair of BSs based on the metric and linkage criterion until all BSs have been clustered. The complete steps of the algorithm are provided in Algorithm~\ref{alg1:Agglomerative_algo}. The inputs are the locations of the BSs $\mathbf{X}_{i}, \forall i \in \{1,\ldots,K\}$ and their NREs $E_{i}(n)$, $\forall i \in \{1,\ldots,K\}, n \in \{1,\ldots,N\}$. The first step is to classify the BSs that are energy deficient on average for the duration of $N$ time slots using the expression $\textbf{S1}(i)$. This is done by evaluating the average probability of a negative NRE of each BS and comparing it with the pre-defined high probability threshold $\phi_{h}$.
We initialize a set $\mathcal{E}^{-}$ containing the indices of all energy deficient BSs, i.e., $\mathcal{E}^{-}=\{i \in \{1,\ldots,K\} : \textbf{S1}(i)\}$ . We also define a BS association matrix $\mathbf{A}$ of size $K \times K$ whose elements indicate the binary connection status between the BSs, i.e., $A_{ij} = 1$ if BS $i$ is connected to BS $j$ and vice versa. Note that $A_{ii} = 0, \forall i \in \{1,\ldots,K\}$, since it is meaningless to connect a BS with itself. From the set $\mathcal{E}^{-}$, we select the most energy deficient BS, denoted by $\hat{i}$, determined by using the expression $\textbf{S2}(i)$. Then, we identify the set of BSs that are within a distance of $r$ km from the selected BS, and denote it as $\mathcal{N}_{\hat{i}}$. This is because physical power lines are restricted by the model to be installed only for distances less than $r$ km. If all neighbouring BSs are energy deficient (checked using the condition in line~\ref{check_all_deficient} of Algorithm~\ref{alg1:Agglomerative_algo}), then energy sharing is not possible. Hence, we remove the selected BS from the set $\mathcal{E}^{-}$ as shown in line~\ref{remove_BS_1} and proceed to the next most energy deficient BS. However, if there is at least one neighbouring BS with surplus energy, we compute the metrics denoted by \textbf{S4}$(\hat{i},j)$ from~\eqref{wt_avg} and~\eqref{wt_stochastic}, where $j$ is the index of the neighbour BS with surplus energy.
The linkage criterion is to maximize the metric. Hence, we select the neighbour BS $\hat{j}$ such that the metric in \textbf{S4}$(\hat{i},\hat{j})$ is the maximum one. The association matrix is updated to reflect that BS $\hat{i}$ and $\hat{j}$ are connected, i.e., $A_{\hat{i}\hat{j}} = A_{\hat{j}\hat{i}} = 1$.

The linked BSs may still be able to participate in further energy sharing opportunities. Hence, their energy status needs to be updated after the link is established between them. We provide the update procedure for the AEA and SEA metrics in Algorithm~\ref{alg1:update}. In the AEA case, after the link is established and energy is transferred by the energy donating BS to the energy deficient one, there are two possibilities: either (i) the donor BS is still left with some energy to be shared with other BSs, or (ii) it has donated all its energy to the energy deficient BS. In the case of (i), i.e., $\mathcal{M}_{\hat{i}\hat{j}} > 0$, we update the average NRE of the donor BS, i.e., $E_{\hat{j}} = \mathcal{M}_{\hat{i}\hat{j}}$ and remove the energy deficient BS from the set $\mathcal{E}^{-}$ indicating that the energy requirements are completely met. On the other hand, in the case of (ii), i.e., $\mathcal{M}_{\hat{i}\hat{j}} \leq 0$, we set the average NRE of the donor BS to zero and replace the average NRE of the energy deficient BS to the remaining deficiency, i.e., $E_{\hat{i}} = \mathcal{M}_{\hat{i}\hat{j}}$. Note that the energy deficient BS has not been removed from the set $\mathcal{E}^{-}$ since its remaining energy deficiency might be supported by another BS in the next iterations. On the other hand, for the SEA metric, a similar course of action is taken by updating the NREs for each time slot. This has a direct impact on the evaluation of probability that classifies a BS as energy deficient as well as the SEA metric. The repeat loop given by lines~\ref{repeat_start} to~\ref{repeat_end}~in~Algorithm~\ref{alg1:Agglomerative_algo} is then repeated until the set $\mathcal{E}^{-}$ becomes empty, i.e., there is no further energy sharing possible via physical links. The final association matrix $\mathbf{A}$ defines the physical connections among~the~BSs.


\begin{algorithm}[t]
\small
\renewcommand{\thealgorithm}{2}
\caption{Divisive Clustering}
\label{divisive_algorithm}
\begin{algorithmic}[1]
\Require BS locations $\mathbf{X}_{i}, \forall i \in \{1,\ldots,K\}$ and NRE $E_{i}(n), \forall i \in \{1,\ldots,K\}, n \in \{1,\ldots,N\}$.
\State Initialize the BS association matrix $\mathbf{A}^{\prime} = \{ (A^{\prime}_{ij}) : A^{\prime}_{ij} = 0 \ \forall i = j, A^{\prime}_{ij} = 1 \ \forall i \neq j\}$.
\State Define the weight matrix, denoted by \textbf{S4}$(i,j), \forall i,j \in \{1,\ldots,K\}$, that assigns weights to the entries in the association matrix based on the AEA and SEA metrics.
\State Find the maximum spanning tree from the weighted graph represented by \textbf{S4}$(i,j), \forall i,j \in \{1,\ldots,K\}$ to obtain the new association matrix $\mathbf{A}$.
\For{$\{i,j \in \{1,\ldots,K \} : A_{ij} = 1\}$}
\If{$\|\mathbf{X}_{i} - \mathbf{X}_{j}\|_{2} > r$,}
\State Remove the link between BS $i$ and $j$, i.e., $A_{ij} = 0$.
\EndIf
\EndFor
\end{algorithmic}
\end{algorithm}

\subsubsection{Divisive Approach}

In this approach, we begin with a mesh network where all BSs are assumed to be connected to each other and then eliminate the redundant and/or infeasible links to obtain the final set of connections among the BSs. The required inputs of divisive algorithm are the locations of the BSs and their associated NREs. We initialize the BS association matrix such that all BSs are connected to each other, i.e., $\mathbf{A}^{\prime} = \{ (A^{\prime}_{ij})  : A^{\prime}_{ij} = 0 \ \forall i = j, A^{\prime}_{ij} = 1 \ \forall i \neq j\}$. Corresponding to each link in the association matrix, we assign a weight given by the matrix expressed as \textbf{S4}$(i,j)$ in Table~\ref{table_expressions}. Using the weight matrix, we construct a maximum spanning tree (MST) from the weighted graph defined by the association matrix $\mathbf{A}^{\prime}$. The MST connects all BSs with the maximum cost of the weighted edges. In other words, it defines the physical links that can be used to transfer energy among the BSs. Several greedy algorithms are available in the literature to convert a connected graph to an MST such as Prim's algorithm and Kruskal's algorithm~\cite{algorithms}. In our problem, the MST is most likely to be unique since the metrics are based on the net energy and distance and there is negligible probability of any two paths having exactly the same cost. Once the MST is obtained, we search for any links that are infeasible due to lack of energy affinity or distance limitations and prune the tree accordingly, i.e., we remove any link in the tree for which $\| \mathbf{X}_{i} - \mathbf{X}_{j} \|_{2} > r$. The association matrix after pruning the tree is denoted by $\mathbf{A}$ and contains the final set of clusters in the graph. The complete sequence of steps is provided in Algorithm~\ref{divisive_algorithm}. 

In summary, two different approaches for obtaining the best combination of physical energy sharing links have been presented based on the NRE available at each BS. The first approach relies on average NRE statistics to obtain the combination of links that are expected to achieve high utilization. The second approach considers the average statistics as well as the variability in the NRE while connecting BSs. This can lead to even higher utilization of established links since the energy sharing status of the links during each time slot is taken into consideration.

\vspace{-0.0cm}
\section{Day-to-Day Cost Minimization} \label{sec_day_to_day}

After determining and installing physical connections among BSs, we develop an optimized energy management framework that minimizes the day-to-day net energy cost of the cellular operator by reducing the electricity procurement from SG. The framework decides the optimal amount of energy to be obtained by each BS from the different available options in order to yield the least cost. The net energy cost of a single BS, denoted by $\Psi_{i}(n)$, depends on the energy transactions as well as the respective prices of the energy sources. It can be obtained by subtracting the total revenue earned by selling extra RE (to SG and to other BSs) from the total expense of purchasing additional energy required (from SG and other BSs) as follows:
\begin{equation}\label{cost_energy}
\small
  \Psi_{i}(n) = c^{\text{g}}(n) q_{i}^{\text{g}}(n) + c^{\text{b}}(n) q_{i}^{\text{b}}(n) - c^{\text{s}}(n) q_{i}^{\text{s}}(n) - c^{\text{e}}(n) q_{i}^{\text{e}}(n).
\end{equation}
To determine the amounts of procured and supplied energies by a BS, we require information about RE generation assuming perfect knowledge of the traffic behaviour. This is crucial since it can have a serious impact on the optimization decisions. As an example, let us assume that during any time slot, if it is known that the RE generation in future time slots will be low, then the optimization framework will be inclined towards conserving the use of energy from the battery in the current time slot and saving it for the future and vice versa. However, in reality the RE generation is random and only prediction models based on historical data are available~\cite{solar_generation}. In this paper, we refer to the availability of these models as having \emph{partial knowledge} about future RE generation. This means we only have access to the statistics of RE generation and not the actual realizations. For bench-marking, we also consider the cases where we have \emph{perfect knowledge}, i.e., full access to all future realizations of the RE in an $N-$slot window, and \emph{zero knowledge}, i.e., access to only the current realization of the generated RE. For the sake of simple presentation, we first present the zero knowledge case followed by the perfect knowledge and partial knowledge cases.

\vspace{-0.0cm}
\subsection{Zero Knowledge}\vspace{-0.0cm}
In the first case, we assume that each BS has access to the current realization of RE generation but is completely unaware of the RE generation in future time slots. Therefore, the optimization decisions during each time slot $n$ are made in real-time and cannot be pre-planned. The $N$ optimization problems in this case are solved at each time slot $n$ and expressed as follows:

\vspace{-0.0in}
\small
\begin{align}
& \underset{ \mathbf{Q}^{\text{Z}}(n)\geq 0}{\text{minimize}}
  \sum \limits_{i=1}^{K} \Psi_{i}(n), \label{Obj1}\\
& \text{subject to} \hspace{0.2cm}
  q_{i}^{\text{g}}(n) + q_{i}^{\text{b}}(n) + q_{i}^{\beta}(n) + \sum \limits^{K}_{j=1, j\neq i} A_{ij} q^{\leftarrow}_{ij}(n) = C_{i}(n), \notag \\
& \hspace{5.5cm}\ \forall \ i \in \{1, \ldots,K \}, \label{const1}\\
& 0 \leq B_{i}(n) \leq B_{\max}, \ \ \forall i = 1,\ldots,K, \label{const2}\\
& \sum \limits_{i=1}^{K} q_{i}^{\text{b}}(n) = \sum \limits_{i=1}^{K} q_{i}^{\text{s}}(n),\label{const3} \\
& A_{ij}q^{\leftarrow}_{ij}(n) \leq  A_{ji}(q^{\rightarrow}_{ji}(n) -  E_{\text{loss}}(q^{\rightarrow}_{ji}(n), \|\mathbf{X}_{i} - \mathbf{X}_{j}\|_{2}) ), \notag \\
 &&& \hspace{1.8cm}\forall i \in \{1, \ldots,K\},j \in \{1, \ldots,K\}, \label{const4}
\end{align}\normalsize
where $\mathbf{Q}^{\text{Z}}(n)$ is the set of optimization variables $\{\mathbf{q}^{\text{g}}(n), \mathbf{q}^{\text{b}}(n), \mathbf{q}^{\text{s}}(n), \mathbf{q}^{\beta}(n), \mathbf{Q}^{\leftarrow}(n), \mathbf{Q}^{\rightarrow}(n)  \}$. The variables are vectors of the form $\mathbf{q}^{x}(n) = [q_{1}^{x}(n), q_{2}^{x}(n), \ldots, q_{K}^{x}(n) ]^{T}$ with $x\in\{\text{g}, \text{b}, \text{s}, \beta\}$, the matrix $\mathbf{Q}^{\leftarrow}(n) = [ q^{\leftarrow}_{ij}(n)] \in \mathbb{R}^{K \times K}$, and the matrix $\mathbf{Q}^{\rightarrow}(n) = [ q^{\rightarrow}_{ij}(n)] \in \mathbb{R}^{K \times K}$. Note that the matrices $\mathbf{Q}^{\rightarrow}$ and $\mathbf{Q}^{\leftarrow}$ are sparse as they depend on the non-zero entries in the matrix $\textbf{A}$.
The objective function in \eqref{Obj1} is the sum of the net energy cost of each BS per time slot. The constraint in \eqref{const1} ensures that the amount of energy obtained from different sources, i.e., SG, battery, and other BSs via physical and virtual connections, meets the energy requirements of the BSs and hence, guarantees a seamless operation. The constraint in \eqref{const2} imposes limits on the charge levels of the batteries at the BSs. Notice that the expression of $B_{i}(n)$ depends on $q_{i}^{\text{e}}(n)$ in addition to the optimization variables. Since there is no prior information about future realizations of RE, the optimization framework will tend to sell remaining energy in the battery, at each time slot, as extra energy. However, it may be worthwhile to store energy to be used in the next time slots to avoid purchasing energy. To control this behaviour, we introduce a parameter, denoted by $B_{th}$, representing the battery threshold level below which the framework will not sell energy back to SG. This is incorporated by defining the extra energy $q^{\text{e}}_{i}(n)$ as follows:
\begin{align}
q_{i}^{\text{e}}(n) = \left[B_{i}(n-1) + \alpha_{i}(n) \right.&\left.- q_{i}^{\beta}(n) - q_{i}^{\text{s}}(n) - \right. \notag \\ &\left. \sum \limits^{K}_{j=1} A_{ij}q^{\rightarrow}_{ij}(n) - B_{\text{th}}\right]^{+},
\end{align}
where $[x]^{+}$ is defined as $\max(0,x)$. Next, the constraint in \eqref{const3}, which is related to the energy sharing via SG, ensures that the total energy bought by the BSs is equal to the total energy sold by other BSs. Finally, the constraint in \eqref{const4} relates to the energy sharing via physical connections and forces the energy obtained by the BS to be less than the energy supplied after subtracting the losses. \textcolor{black} {The role of the BS association matrix $\mathbf{A}$ is to select only the constraints related to active links between BSs after the clustering process}.
Notice that the objective and constraints, except~\eqref{const4}, are linear. The constraint in \eqref{const4} involves a convex quadratic term due to the loss function given in \eqref{P_loss}. Hence, the problem is convex and can be efficiently solved by off-the-shelf solvers such as CVX~\cite{cvx}.

\vspace{-0.0cm}
\subsection{Perfect Knowledge}\vspace{-0.0cm}
\textcolor{black}{In the perfect knowledge case, we assume that, $\forall \ i \in \{1, \ldots,K\},n \in \{1, \ldots, N \}$, $\alpha_{i}(n)$ is predicted perfectly. Other information, i.e., price and traffic variations, are also known \emph{a priori} to the BS. Although this may not be true in practice, however, the approach is still practical for extremely reliable RE estimates. Moreover, the perfect knowledge case serves as a useful benchmark for comparison with other strategies}. In this case, the optimized energy decisions can be obtained by solving a single optimization problem minimizing the sum of energy cost over the time slots expressed as follows:

\vspace{-0.0in}
\small
\begin{align}
& \underset{ \mathbf{Q}^{\text{P}}\geq 0 }{\text{minimize}} \sum \limits_{n=1}^{N} \sum \limits_{i=1}^{K} \Psi_{i}(n), \label{Obj2}\\
& \text{subject to } \;\; q_{i}^{\text{g}}(n) + q_{i}^{\text{b}}(n) + q_{i}^{\beta}(n) + \sum \limits^{K}_{j=1} A_{ij}q^{\leftarrow}_{ij}(n) = C_{i}(n), \notag \\
&\hspace{3.4cm} \forall i \in \{1, \ldots, K \}, n \in \{1,\ldots,N\},\label{const21}\\
& 0 \leq B_{i}(n) \leq B_{\max},\ \forall \ i \in \{1,\ldots,K\},n \in \{1,\ldots,N\},\label{const22}\\
&\sum \limits_{i=1}^{K} q_{i}^{\text{b}}(n) = \sum \limits_{i=1}^{K} q_{i}^{\text{s}}(n), \forall \ n \in \{1,\ldots,N\},\label{const23} \\
& A_{ij}q^{\leftarrow}_{ij}(n) \leq  A_{ji}(q^{\rightarrow}_{ji}(n) -  E_{\text{loss}}(q^{\rightarrow}_{ji}(n), \|\mathbf{X}_{i} - \mathbf{X}_{j}\|_{2}) ), \notag \\
 & \hspace{1cm}\forall i \in \{1, \ldots,K\},j \in \{1,\ldots,K\}, n\in \{1,\ldots,N\}, \label{const24}
\end{align}\normalsize
where $\mathbf{Q}^{\text{P}}$ is the set of optimization variables $\{ \mathbf{q}^{\text{g}}(n), \mathbf{q}^{\text{b}}(n), \mathbf{q}^{\text{s}}(n), \mathbf{q}^{\beta}(n), \mathbf{q}^{\text{e}}(n), \mathbf{Q}^{\leftarrow}(n), \mathbf{Q}^{\rightarrow}(n),\\ \forall n \in \{1, \ldots, N \}  \}$.
The objective in \eqref{Obj2} is now to minimize the net cost of energy of all BSs over all time slots. The constraint in \eqref{const21} ensures that the total amounts of energies obtained from SG, other BSs via physical and virtual connections, and the battery is equal to the energy consumption of the BS. The constraint in \eqref{const22} puts essential limits on the charge level of the BS batteries at all time slots. Since there is perfect knowledge of all input variables in a particular time window, the battery level can now be re-defined for all time slots as follows:
\vspace{-0.0cm}

\begin{align}
  B_{i}(n) =&  \ B_{0} + \sum \limits_{k=1}^{n} \alpha_{i}(k) - \sum \limits_{k=1}^{n} q^{\beta}_{i}(k)  - \sum \limits_{k=1}^{n} q_{i}^{\text{s}}(k)  - \notag \\& \sum \limits_{k=1}^{n} \sum \limits_{j = 1}^{K} A_{ij} q^{\rightarrow}_{ij}(k)  - \sum \limits_{j=1}^{n} q_{i}^{\text{e}}(j), \forall \ i\in\{1,\ldots,K\}, \notag \\ & \hspace{4cm} n\in\{1,\ldots,N\}.
\end{align}\normalsize

The constraints in \cref{const23,const24} are similar to the zero knowledge case.
It is clear that the problem has a linear objective function and a set of linear constraints except \eqref{const24}. However, it is easy to show that \eqref{const24} is convex and hence, the problem is a convex optimization problem that can be efficiently solved by off-the-shelf solvers such as CVX~\cite{cvx}. \textcolor{black}{Note that the complexity of the problem in the perfect knowledge case is significantly higher than that of the zero knowledge case. This is because of the higher number of optimization variables (i.e., at maximum $(5K + 2K^{2})N$ in case of a mesh network) as compared to the zero knowledge case which has at maximum $4K + 2K^{2}$ variables excluding the variables $q_{i}^{\text{e}}(n)$ that are not part of the optimization.
However, this does not pose a serious concern for the operator since the optimization needs to be carried out only once for the period of $N$ time slots as opposed to the zero knowledge case which needs to be executed after each time slot.}

\vspace{-0.0cm}
\subsection{Partial Knowledge}\vspace{-0.0cm}
\normalsize
In this case, we assume that the BSs only have partial information about the RE generation in future time slots. This means that a portion of the predicted RE generation is uncertain. \textcolor{black}{Therefore, we model the RE generation matrix $\boldsymbol{\alpha}$ with elements $\alpha_{i}(n), \forall \ i \in \{1,\ldots,K\},n \in \{ 1,\ldots,N\}$ as $\boldsymbol{\alpha} = \bar{\boldsymbol{\alpha}} + \tilde{\boldsymbol{\alpha}},$ where $\bar{\boldsymbol{\alpha}}$ is the deterministic portion of the RE generated, estimated from historical data, and $\boldsymbol{\tilde{\alpha}}$ is a $K\times N$ matrix of random variables representing the stochastic portion.} 

In order to cater for the uncertainty in $\boldsymbol{\alpha}$, we re-formulate the problem in \cref{Obj2,const21,const22,const23,const24} as a stochastic optimization problem~\cite{stochastic_programming2}. Assuming that we know one of the output variables, we can optimize other variables for any given value of $\boldsymbol{\alpha}$. However, the decision needs to be updated once the actual realization of $\boldsymbol{\alpha}$ has been obtained. We choose to fix feasible values of the variables $\mathbf{q}^{\text{g}}(n), \forall \ n \in \{ 1, \ldots,N\}$ since they are associated with the price $c^{\text{g}}(n)$ that is beyond the control of the BS. Hence, fixing the vector $\mathbf{q}^{\text{g}}(n), \forall \ n \in \{ 1, \ldots,N\}$ allows us to compute the best combination of other variables provided that $\boldsymbol{\alpha}$ is known. The problem can be written as a two-stage recourse problem as follows:

\vspace{-0.0in}
\small
\begin{align}
& \underset{\mathbf{q}^{\text{g}}(n) > 0, \ \forall n \in \{1,\ldots,N\}}{\text{minimize}} \ \ \
 \sum \limits_{n=1}^{N} \sum \limits_{i=1}^{K} c^{\text{g}}(n) q_{i}^{\text{g}}(n) + \mathbb{E}_{\boldsymbol{\alpha}} [ \Psi^{\star} ], \label{Obj3}
\end{align}

\normalsize
\noindent where $\mathbb{E}_{\boldsymbol{\alpha}}[.]$ represents the expectation function with respect to $\boldsymbol{\alpha}$ and $\Psi^{\star}$ is obtained as follows:
\small
\begin{align}
& \Psi^{\star} = \underset{\mathbf{\tilde{Q}} > 0}{\text{minimize}} \ \ f(\mathbf{\tilde{Q}}), \label{Obj4}\\
& \text{subject to } \;\; q_{i}^{\text{g}}(n) + q_{i}^{\text{b}}(n) + q_{i}^{\beta}(n) + \sum \limits^{K}_{j=1} A_{ij} q^{\leftarrow}_{ij}(n) = C_{i}(n), \notag \\
& \hspace{3.3cm} \forall i \in \{1, \ldots, K \}, n \in \{1,\ldots,N\}, \label{const41}\\
&\sum \limits_{k=1}^{n} q_{i}^{\beta} + \sum \limits_{k=1}^{n} q_{i}^{\text{s}} + \sum \limits_{k=1}^{n}  \sum \limits_{j=1}^{K} A_{ij} q^{\rightarrow}_{ij}(k) \leq B_{0} + \sum \limits_{k=1}^{n} \alpha_{i}(k), \notag \\
& \hspace{3.3cm}\forall \ i \in \{1,\ldots,K\}, n \in \{1, \ldots,N\},   \label{const42}\\
& -\sum \limits_{k=1}^{n} q_{i}^{\beta} -  \sum \limits_{k=1}^{n} q_{i}^{\text{s}} - \sum \limits_{k=1}^{n}  \sum \limits_{j=1}^{K} A_{ij} q^{\rightarrow}_{ij}(k) \leq B_{\max}-B_{0} - \notag \\
& \hspace{1.3cm} \sum \limits_{k=1}^{n} \alpha_{i}(k), \forall \ i \in \{1,\ldots,K\}, n \in \{1, \ldots,N\},\label{const43} \\
& \sum \limits_{i=1}^{K} q_{i}^{\text{b}}(n) = \sum \limits_{i=1}^{K} q_{i}^{\text{s}}(n), \forall \ n \in \{1,\ldots,N\},\label{const44}\\
& A_{ij}q^{\leftarrow}_{ij} \leq  A_{ji}(q^{\rightarrow}_{ji} -  E_{\text{loss}}(q^{\rightarrow}_{ji}, \|\mathbf{X}_{i} - \mathbf{X}_{j}\|_{2}) ), \notag \\
& \hspace{3.3cm}\forall i \in \{1, \ldots,K\},j \in \{1,\ldots,K\} \label{const45},
\end{align}\normalsize

\noindent where $\mathbf{\tilde{Q}}$ represents the set of second stage optimization variables $\{\mathbf{q}^{\text{b}}(n), \mathbf{q}^{\text{s}}(n), \mathbf{q}^{\beta}(n), \mathbf{q}^{\text{e}}(n), \\ \mathbf{Q}^{\leftarrow}(n), \mathbf{Q}^{\rightarrow}(n)  \}$.
The function $f(\mathbf{\tilde{Q}})$ is defined as follows:
\begin{equation}
\small
f(\mathbf{\tilde{Q}}) = \sum \limits_{n=1}^{N} \sum \limits_{i=1}^{K} \left(  c^{\text{b}}(n) q_{i}^{\text{b}}(n) - c^{\text{s}}(n) q_{i}^{\text{s}}(n) - c^{\text{e}}(n) q_{i}^{\text{e}}(n)  \right).
\end{equation}
%
If the solution to the second stage problem expressed in \cref{Obj4,const41,const42,const43,const44,const45} is obtained in a closed form, then a tractable solution to the first stage problem in~\cref{Obj3} can be obtained after evaluating the expectation over $\boldsymbol{\alpha}$. However, in most cases, obtaining a closed form solution may either be impossible or results in analytically complicated expressions. Hence, the random variables are often discretized to make the two-stage recourse problem tractable for numerical solvers~\cite{discretization}. In our case, we assume that the random variables $\tilde{\alpha}_{i}(n)$ have been discretized to take a set of $M$ possible values. The resulting set of $\mathcal{M} = M^{KN}$ possibilities of the matrix $\boldsymbol{\alpha}$ is denoted by $\Omega$. We also assume that the probability mass function (pmf) of $\boldsymbol{\tilde{\alpha}}$ can be estimated and hence, the pmf of $\boldsymbol{\alpha}$, denoted by $P_{m}$ for $m = 1, \ldots, \mathcal{M}$, can directly be obtained. Consequently, the two-stage problem can be formulated as one large convex optimization problem, also known as the deterministic equivalent of the original problem, which is expressed as follows:

\vspace{-0.0in}
\small
\begin{align}
&\underset{\mathbf{Q}^{\text{g}}}{\text{minimize}}\,\,\sum \limits_{n=1}^{N} \sum \limits_{i=1}^{K} c^{\text{g}}(n) q_{i}^{\text{g}}(n) + \mathbb{E}_{\alpha} [f(\mathbf{\tilde{Q}})] , \label{Obj5}\\
& \text{subject to}\,\,q_{i}^{\text{g}}(n) + q_{i,m}^{\text{b}}(n) + q_{i,m}^{\beta}(n) + \sum \limits^{K}_{j=1} q^{\leftarrow}_{ij,m}(n) = C_{i}(n), \notag \\
 & \hspace{1cm} \forall i \in \{1, \ldots, K \}, n \in \{1,\ldots,N\},  m \in \{ 1, \ldots, \mathcal{M}\}, \label{const51}\\
&\sum \limits_{k=1}^{n} q_{i,m}^{\beta} + \sum \limits_{k=1}^{n} q_{i,m}^{\text{s}} + \sum \limits_{k=1}^{n}  \sum \limits_{j=1}^{K} A_{ij} q^{\rightarrow}_{ij,m}(k) \leq B_{0} + \sum \limits_{k=1}^{n} \alpha_{i,m}(k), \notag \\
& \hspace{1cm} \forall \ i \in \{1,\ldots,K\}, n \in \{1, \ldots,N\}, m \in \{ 1, \ldots, \mathcal{M}\},  \label{const52}\\
&-\sum \limits_{k=1}^{n} q_{i,m}^{\beta} -  \sum \limits_{k=1}^{n} q_{i,m}^{\text{s}} - \sum \limits_{k=1}^{n}  \sum \limits_{j=1}^{K} A_{ij} q^{\rightarrow}_{ij,m}(k) \leq B_{\max}-B_{0} - \notag \\
& \hspace{0cm} \sum \limits_{k=1}^{n} \alpha_{i,m}(k), \, \forall \ i \in \{1,\ldots,K\}, n \in \{1, \ldots,N\}, m \in \{ 1, \ldots, \mathcal{M}\},\label{const53} \\
&\sum \limits_{i=1}^{K} q_{i,m}^{\text{b}}(n) = \sum \limits_{i=1}^{K} q_{i,m}^{\text{s}}(n), \ \forall \ n \in \{1,\ldots,N\}, m =\{1, \ldots, \mathcal{M}\},\label{const54}\ \\
&A_{ij}q^{\leftarrow}_{ij,m} \leq  A_{ji}(q^{\rightarrow}_{ji,m} -  E_{\text{loss}}(q^{\rightarrow}_{ji,m}, \|\mathbf{X}_{i} - \mathbf{X}_{j}\|_{2}) ), \notag \\
 & \hspace{1cm}\forall i \in \{1, \ldots,K\},j \in \{1,\ldots,K\}, m \in \{ 1, \ldots, \mathcal{M}\},\label{const55}
\end{align}\normalsize
where the subscript $m$ in the variables is added to indicate their values for the $m^{\text{th}}$ possibility of the matrix $\boldsymbol{\alpha}$ and the expectation in the objective is evaluated as $\mathbb{E}_{\alpha} [f(\mathbf{\tilde{Q}})] = \sum \limits_{m = 1}^{\mathcal{M}} P_{m} f(\mathbf{\tilde{Q}}_m)$.
Notice that the number of constraints of the problem exponentially scales with the number of possibilities $M$, the number of BSs $K$, and the number of time slots $N$. The solution of the convex problem \eqref{Obj5}-\eqref{const55} can be obtained by using off-the-shelf solvers such as CVX~\cite{cvx}.

\vspace{-0.0cm}
\section{Numerical Results} \label{sec_results}
In this section, we present the simulation model and investigate the performance of the proposed energy sharing algorithms and metrics in addition to the optimized energy procurement.
\vspace{-0.0in}
\subsection{Simulation Model} \label{sim_parameters}
We consider a $5$ $\times$ $5$ km$^2$ square region in $\mathbb{R}^{2}$ with $K = 20$ cellular BSs placed uniformly according to a realization of the hard core point process with an exclusion distance of $500$ m. This ensures that each pair of BSs is separated by at least the minimum exclusion distance to avoid the case of very closely located BSs that may appear in commonly used Poisson point processes. The BSs are randomly labeled as $\{1,\ldots,K\}$ for ease of referencing. A snapshot of the network realization can be seen in Fig.~\ref{clustering_fig}. Note that the choice of BS locations and other simulation parameters is arbitrary and is used for illustrative purposes only without loss of generality in the framework or the results. We assume a total network operation time of $N = 24$ hours divided into hourly time slots, i.e., $n = 1,2, \ldots, N$. The RE at BSs sites is considered to be generated from solar panels only and is modeled as $\alpha_{i}(n) = \bar{\alpha}_{i}(n) + \tilde{\alpha}_{i}(n)$, where the average RE, $\bar{\alpha}_{i}(n)$, is characterized by the following model~\cite{solar_generation}:
\vspace{-0.0cm}
\begin{equation}\label{alpha_bar}
  \bar{\alpha}_{i}(n) = \frac{\alpha_{i}^{\max} \exp^{-(n - \mu^{\alpha}_{i})^{2}}}{(\sigma^{\alpha}_{i})^2}\tau.
\end{equation}

\normalsize
\textcolor{black}{In this model, $\alpha_{i}^{\max} = \mathcal{A}_{i}\mathcal{I}_{i}\eta$ represents the maximum power generation capacity of BS $i$ where $\mathcal{A}_{i}$ is the surface area of the solar panel, $\mathcal{I}_{i}$ is the peak irradiance and $\eta$ is the energy conversion efficiency. In our simulations, we use the following parameters: $\mathcal{A}_{i} = 1$ m$^2$, $\mathcal{I}_{i} = 1$ kW/m$^{2}$, $\forall i \in \{1,\ldots,K\}$, and $\eta = 20 \%$}. The parameter $\mu_{i}^{\alpha}$ represents the position in time of the peak generation, chosen to be mid-day, i.e., $12$ hrs, $\forall i \in \{1,\ldots,K\}$, while $\sigma_{i}^{\alpha}$ represents the shape width at half maximum of the peak, chosen to be $3$ hrs, $\forall i \in \{1,\ldots,K\}$. The time duration of each slot $\tau$ is $1$ hr. The randomness in RE $\tilde{\alpha}_{i}(n)$ is modeled as a zero mean Gaussian random variable with a standard deviation of $5$ W. For the partial knowledge case, we assume that the random variable ˜$\tilde{\alpha}_{i}(n)$ can only take the discrete values $0.2\bar{\alpha}_{i}(n)$ and $-0.2\bar{\alpha}_{i}(n)$ with equal probabilities. This signifies that the variation in the RE generation is around $20$\% to the mean. Each BS owns a battery to store the generated RE with maximum capacity $B_{\max} = 100$ Wh. The batteries are initially assumed to be fully charged, i.e., $B_{0}= B_{\max}$ Wh, unless otherwise specified. For the zero knowledge case, we set the battery threshold level to $B_{\text{th}} = 0.5 B_{\max}$.

\textcolor{black}{The power consumption of the BSs, which is dependent on the traffic, is chosen to have a bi-modal Gaussian profile over time (i.e., $\sum_{i \in \{a,b\}} \frac{\gamma_i}{\sqrt{2 \pi \sigma_i^C}} \exp(-  \left( \frac{n - \mu_i^C}{\sigma_i^C}\right)^2 )$) with means $\mu_{a}^{C} = 10$ hrs (i.e., $10$ am), $\mu_{b}^{C} = 18$ hrs (i.e., $6$ pm), standard deviations $\sigma_{a}^{C} = \sigma_{b}^{C} = 3$ hrs, and mixing parameters $\gamma_a = 0.6$ and $\gamma_b = 0.4$}. This is consistent with actual cellular traffic measurements in urban areas that show two peak traffic times each day~\cite{traffic_model}. \textcolor{black}{The maximum power consumption of the BSs $\rho^{\max}$ is evaluated as follows: $\rho^{\max} = a (\rho^{\text{tx}}\nu^{\max}) + b$, where $\rho^{\text{tx}}$ is the transmit power per user, $\nu^{\max}$ is the maximum number of users supported by a BS, $a$ is the scaling parameter, and $b$ is the constant power consumption of the BS irrespective of the number of connected users. According to the EARTH model~\cite{EARTH}, the parameters $a = 4.7$ W, $b = 130$ W for macro BSs. We choose the maximum number of subscribers $\nu^{\max} = 50$ and $P^{\text{tx}} = 0.3$ W}. The energy trading prices used in the paper are as follows: the purchasing price per unit from SG is $c^{\text{g}}(n) = 0.8$ MU, the selling price of extra energy to SG is $c^{\text{e}}(n) = 0.2$ MU, the price of buying energy from other BSs via SG is $c^{\text{b}}(n) = 0.6$ MU, and the price for selling energy to other BSs via SG is $c^{\text{s}}(n) = 0.4$ MU. The prices are assumed to be constant for all time slots $n \in \{1,\ldots,N\}$.

For the physical connections among BSs, we assume that multi-core aluminum conductors with cross sectional area of $300$ mm$^{2}$ are used. The specific resistance of such power cables is evaluated to be $0.113$ m$\Omega$/m at $45^{\circ}$C\cite{cable_parameters}.
It is also important here to mention the parameters required for the evaluation of SEA metric in \eqref{wt_stochastic}. In this simulation setting, since $\alpha_{i}(n)$ and $C_{i}(n)$ are Gaussian random variables, the NRE $E_{i}(n) = \alpha_{i}(n) - C_{i}(n)$, is also Gaussian with mean $\mu_{i}^{E} = \mu_{i}^{\alpha} - \mu_{i}^{C}$ and standard deviation $\sigma_{i}^{E} = \sqrt{(\sigma_{i}^{\alpha})^{2} + (\sigma_{i}^{C})^{2}}$. Assuming $E_{i}$ and $E_{j}$ to be independent and denoting $Z_{ij}(n) = E_{i}(n) - E_{j}(n)$, the probability in the SEA metric in \eqref{wt_stochastic} can be evaluated as follows:

\vspace{-0.0in}
\small
\begin{equation}\label{prob_stochastic}
  \mathbb{P}[|E_{i}(n) - E_{j}(n)| > \delta] = 1 - \frac{1}{2} \text{erf}\left(\frac{ \delta - \mu_{ij}^{Z}}{\sigma_{ij}^{Z} \sqrt{2}  }\right) + \frac{1}{2} \text{erf}\left(\frac{- \delta - \mu_{ij}^{Z}}{\sigma_{ij}^{Z} \sqrt{2}  }\right),
\end{equation}
\normalsize

\noindent where $\text{erf}(.)$ denotes the Gaussian error function, $\mu_{ij}^{Z} = \mu_{i}^{E} - \mu_{j}^{E}$, and $\sigma_{ij}^{Z} = \sqrt{(\sigma_{i}^{E})^{2} + (\sigma_{j}^{E})^{2}}$. For the condition statement in \eqref{wt_stochastic}, the probability is calculated as follows:

\vspace{-0.0in}
\small
\begin{align}\label{metric_condition}
    \mathbb{P} [ &\underset{k \in \{i,j\}}{\cap} \{E_{k}(n) > 0\} \ \text{or} \underset{k \in \{i,j\}}{\cap} \{E_{k}(n) < 0\}  ] \notag =\\& \mathbb{P}[\{E_{i} > 0\} \cap \{E_{j} > 0\}] + \mathbb{P}[\{E_{i} < 0\} \cap \{E_{j} < 0\}], \notag \\
    &= \frac{1}{2} \left[ 1 + \text{erf}\left(\frac{- \mu_{i}^{E}}{\sqrt{2}\sigma_{i}^{E}} \right) \text{erf}\left(\frac{- \mu_{j}^{E}}{\sqrt{2} \sigma_{j}^{E}}\right) \right].
\end{align}
\normalsize

The reference parameters for the SEA metric are selected as follows: the reference energy gap $\delta$ is chosen to be $0$ MU (i.e., sharing even for small energy differences), low probability threshold $\phi_{l} = 0.5$, and high probability threshold $\phi_{h} = 0.5$. Note that these parameters are selected to encourage establishment of maximum energy sharing links. If these reference parameters are tightened, the clustering algorithms will become more selective and fewer links will be formed.

\vspace{-0.0cm}
\subsection{Performance of Clustering Algorithms \& Metrics} 
\begin{figure}[t!]
  \centering
  \includegraphics[width=3.5in]{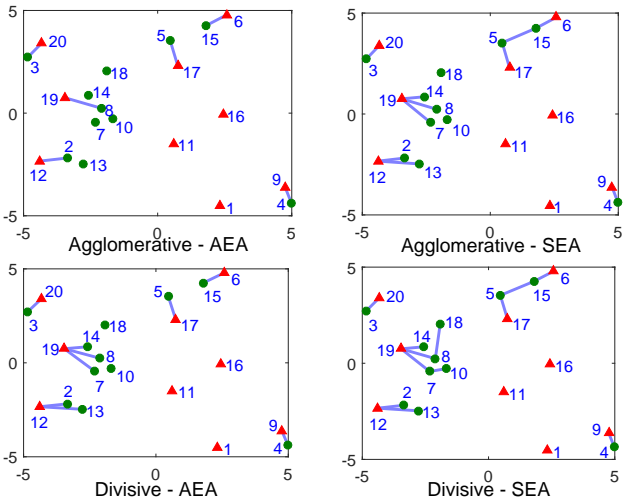}\vspace{-0.0in}
  \caption{Example physical connections using different clustering algorithms and metrics for energy sharing range $r = 2$ km. \textcolor{black}{Red triangles indicate BSs that have negative average NRE and green circles represent the BSs with positive average NRE.}}\label{clustering_fig}\vspace{-0.0in}
\end{figure}

\vspace{-0.0cm}
In this subsection, the behaviour and impact of the two proposed clustering algorithms and the underlying metrics are investigated. Fig.~\ref{clustering_fig} shows the physical links obtained using the agglomerative and divisive clustering algorithms employing both AEA and SEA metrics for the chosen network realization. The red triangles represent the BSs having an energy deficiency on average (i.e., average NRE $<$ 0) while the green circles represent the BSs having surplus energy on average (i.e., average NRE $>$ 0). The blue connecting lines represent the physical power lines installed to share energy among BSs. The figure shows that the links established via the agglomerative approach are a subset of the links established via the divisive approach. Furthermore, the links established via the AEA metric are a subset of the links established via the SEA metric.
Although the divisive algorithm may link more BSs than the agglomerative one for smaller energy sharing ranges, this is not true in general for larger energy sharing ranges because the MST prevents establishment of cliques and loops in the graph to maintain the tree structure. The agglomerative algorithm, on the other hand, has more freedom in establishing links and hence, can lead to a higher number of energy sharing links.

On the other hand, the agglomerative algorithm only links two BSs if the link is highly favourable. Therefore, some links might not be established in the agglomerative algorithm which are present in the divisive algorithm.
\begin{figure*}[h!]
\addtolength{\subfigcapskip}{-0.0in}
\begin{center}
\subfigure[]{\label{3a}\includegraphics[width=2.05in]{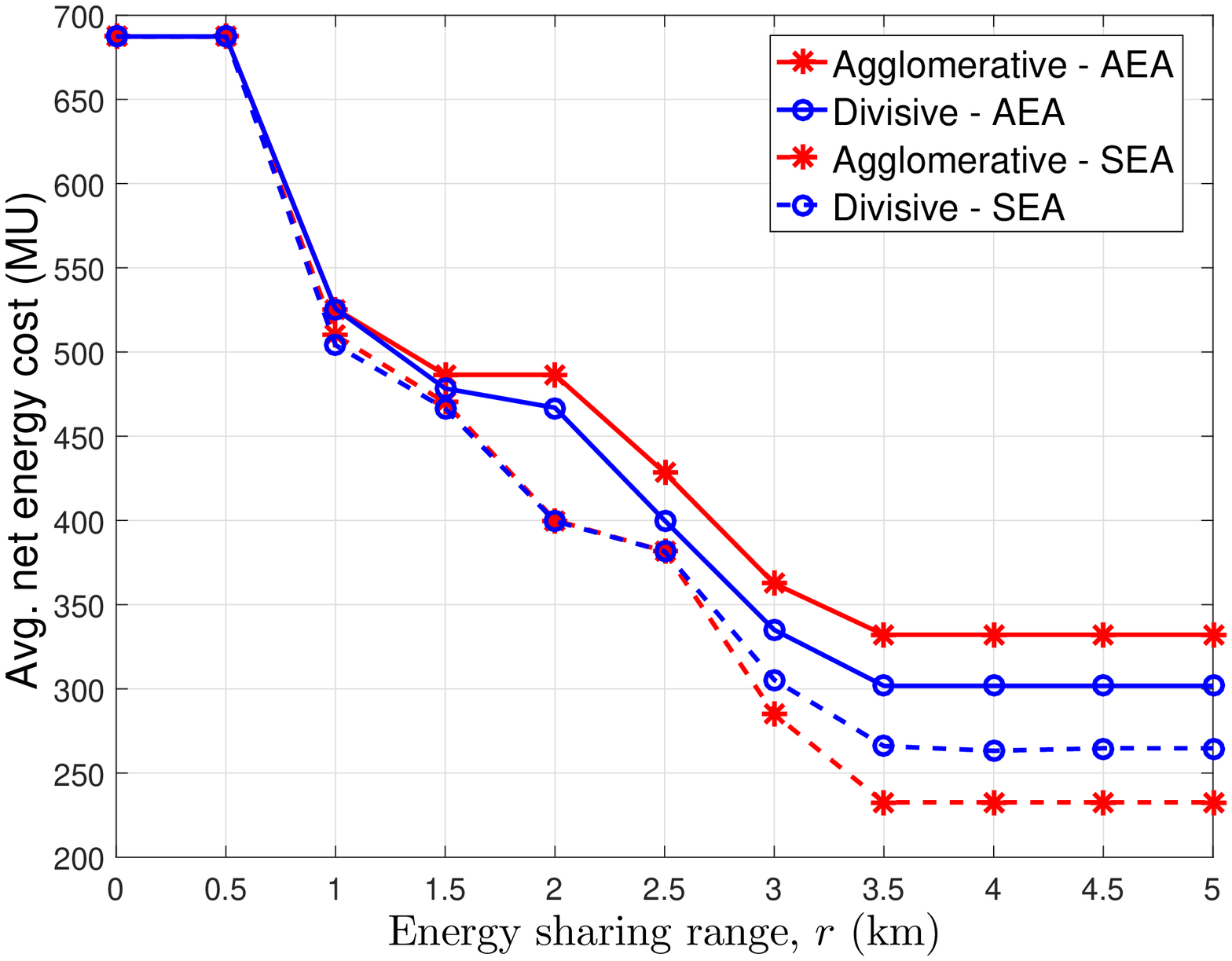}}
\subfigure[]{\label{3b}\includegraphics[width=2.05in]{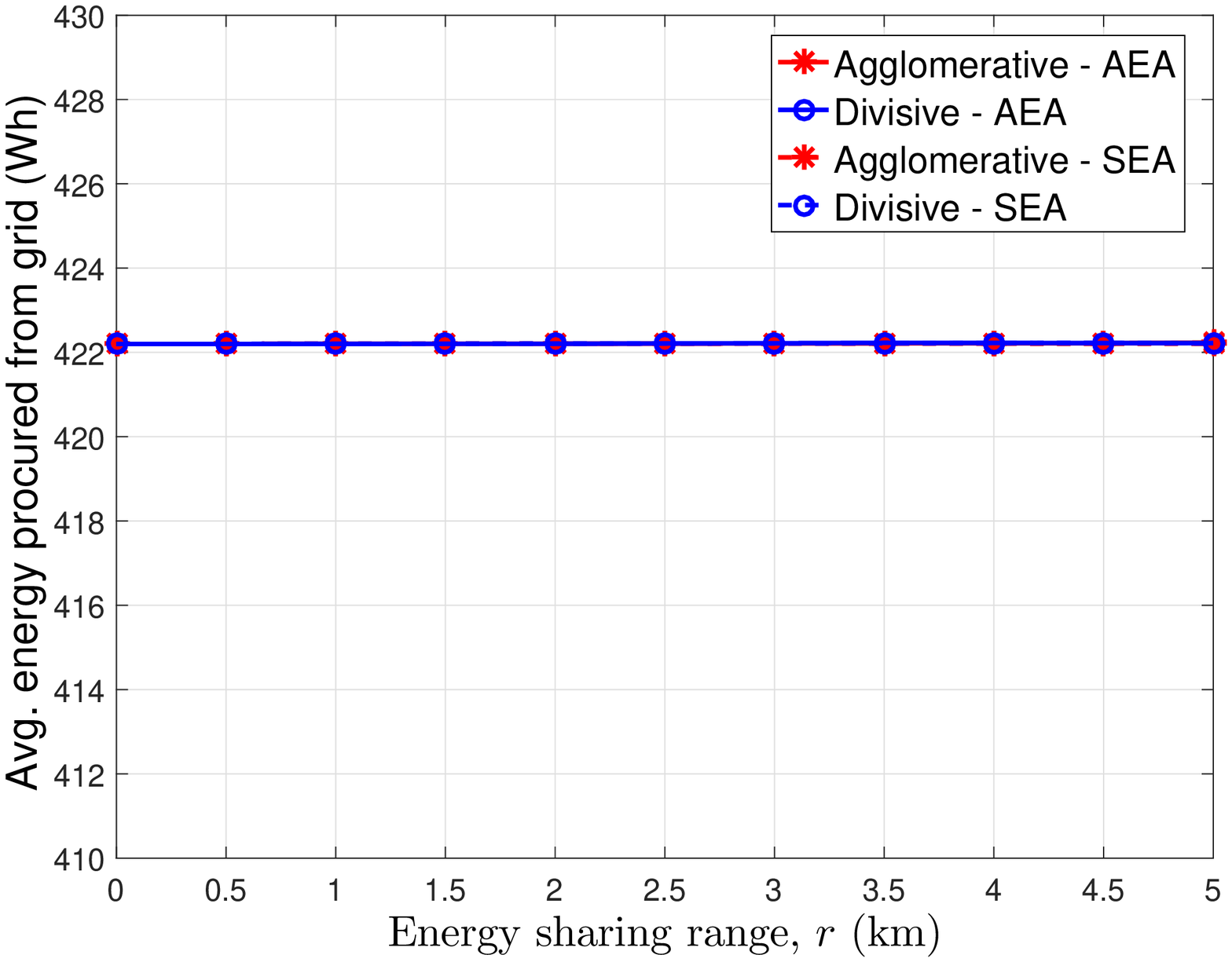}}\
\subfigure[]{\label{3c}\includegraphics[width=2.05in]{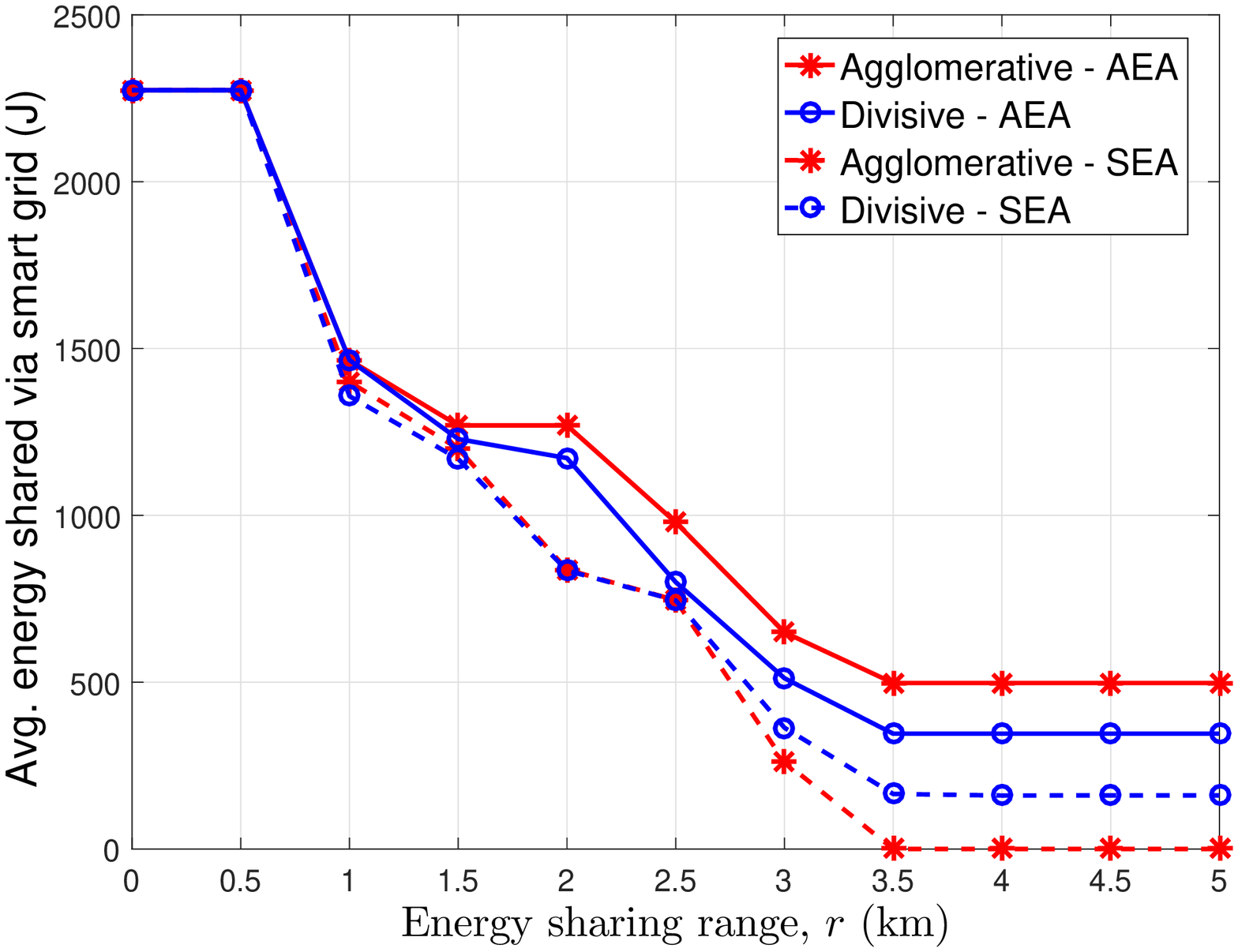}}\\
\subfigure[]{\label{3d}\includegraphics[width=2.05in]{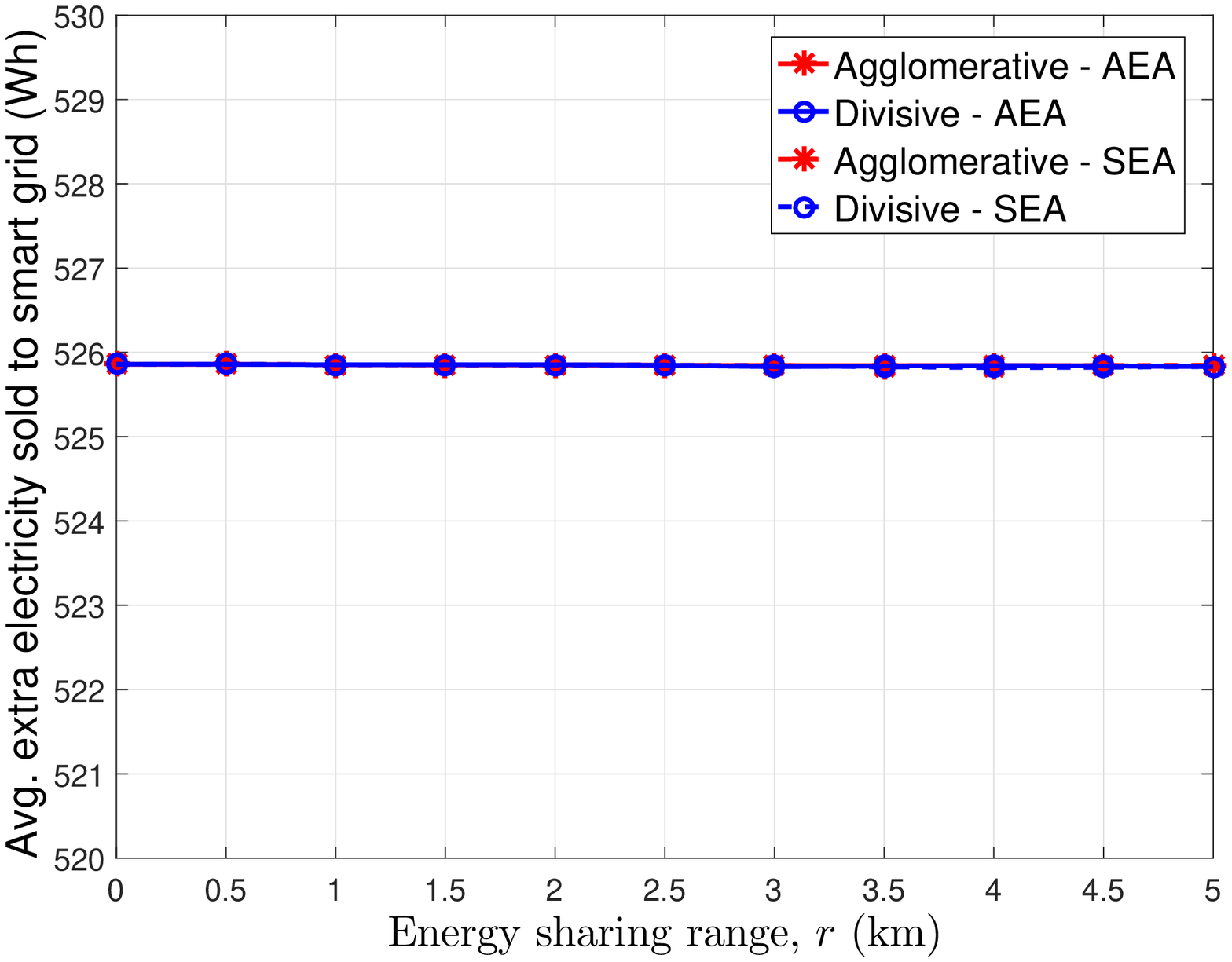}}
\subfigure[]{\label{3e}\includegraphics[width=2.05in]{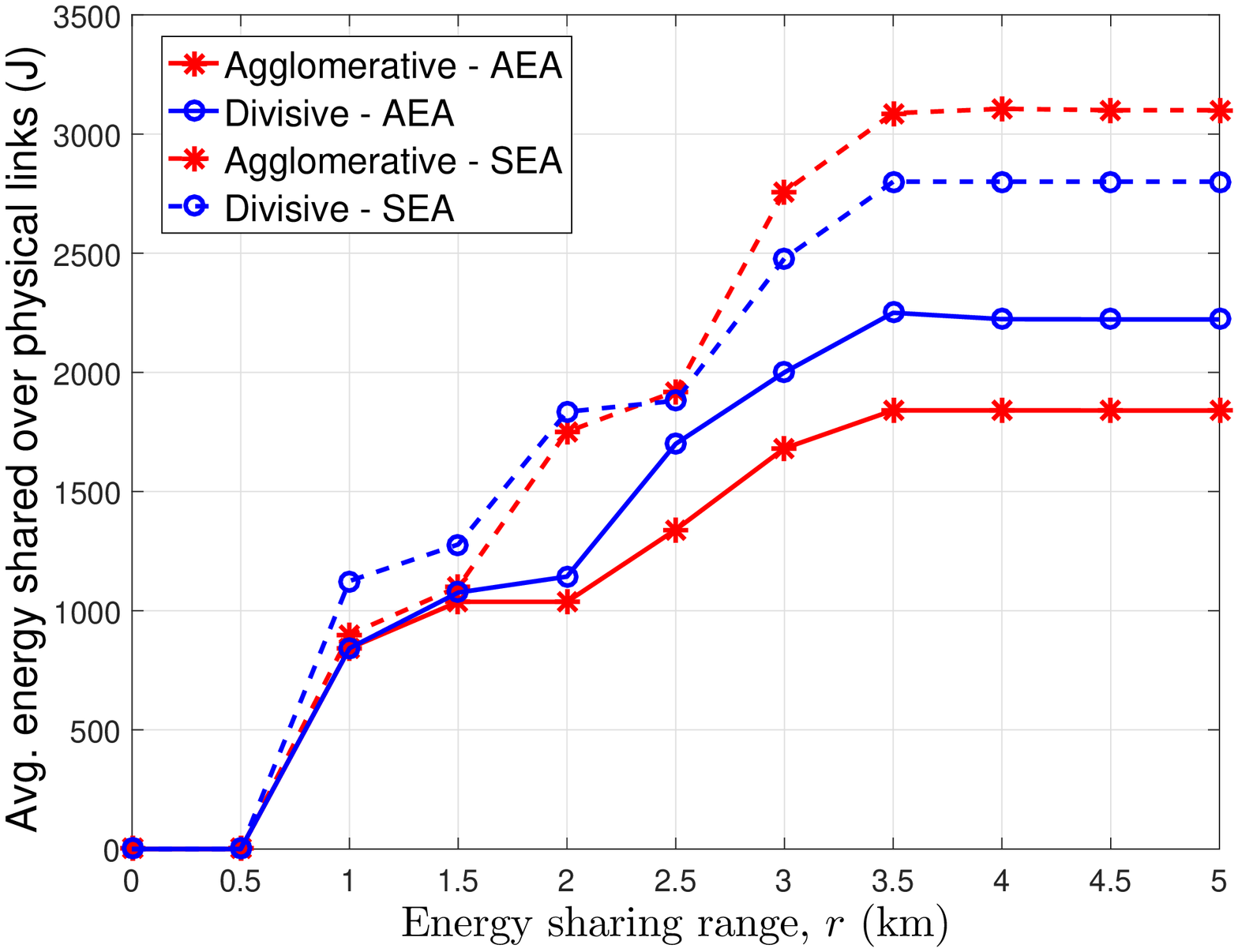}}
\subfigure[]{\label{3f}\includegraphics[width=2.05in]{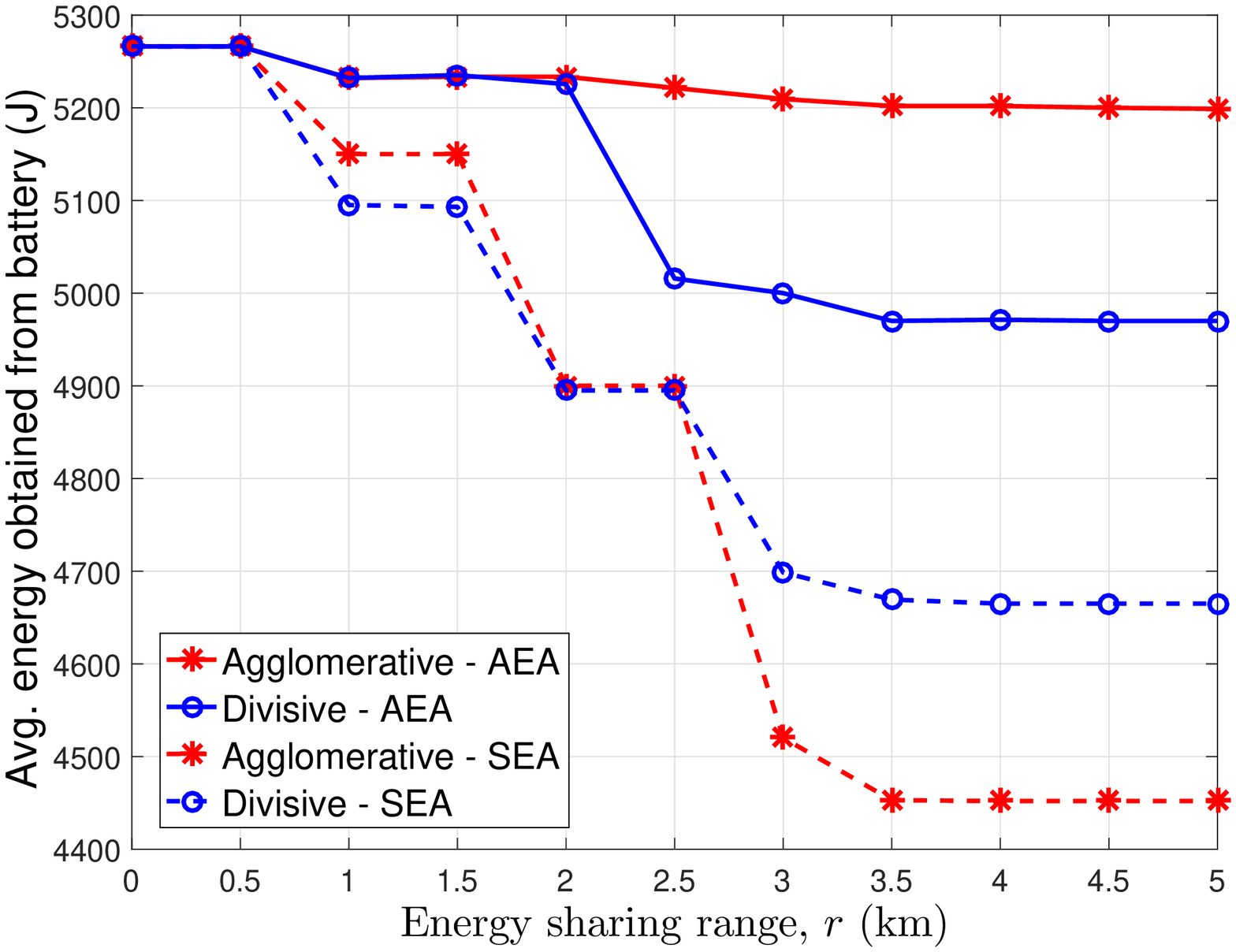}}\\
\subfigure[]{\label{3g}\includegraphics[width=2.05in]{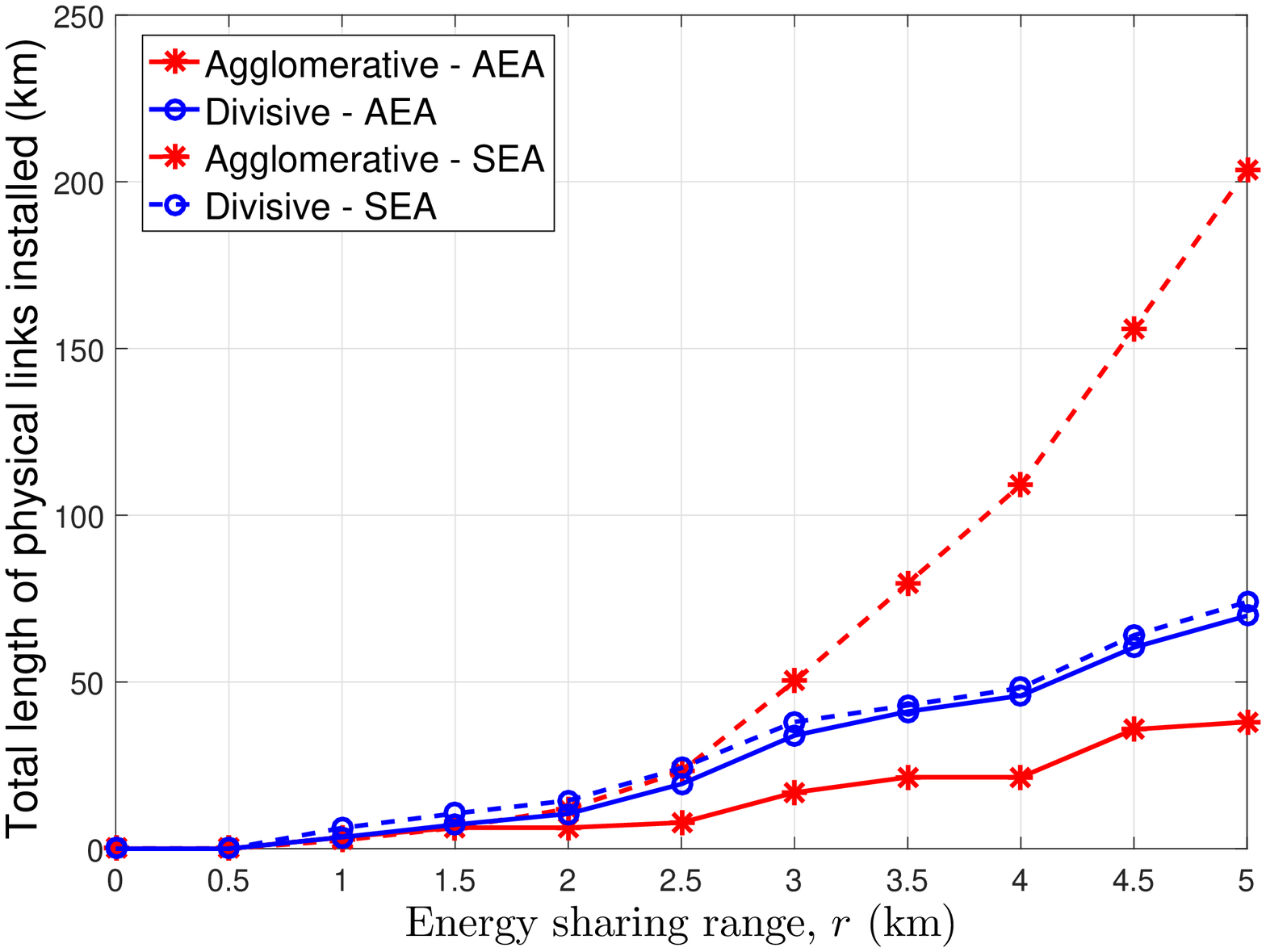}}
\end{center}
\caption{Total cost and associated energy transactions versus energy sharing range; (a) average net energy cost, (b) average electricity procured from SG, (c) average energy shared via SG, (d) average extra electricity sold back to SG, (e) average energy shared over physical links, (f) average energy obtained from the battery, and (g) total length of installed physical links.}
\label{3}
\end{figure*}
The SEA metric is sensitive to the NREs across time slots. Hence, the SEA adds more links between BSs that experience different NREs across time even if they have an overall positive average NRE. Consequently, all connections in the AEA scheme are established between different coloured BSs only, which clearly identifies the energy donating BS and energy receiving BS. Such discrimination is not possible when using the SEA metric because there is a possibility that despite having an overall positive average NRE, the BS might have some time slots of negative NRE. Hence, it may be beneficial to connect it to another BS that has a higher chance of having surplus energy during the day. As an example, consider the link between BSs $5$ and $15$. Both have an overall positive NRE but the link between them may be useful in some cases considering variability in RE generation and consumption pattern of the BS. It is pertinent to mention here that physical links obtained using the SEA metric can be different if the reference parameters $\delta$, $\phi_{l}$, and $\phi_{h}$ are modified. However, in general, the SEA metric is considerate of the variability in NREs of BSs and thus, enables more effective use of energy sharing links as compared to links based on average statistics in AEA metric.

In order to investigate the benefit of the obtained physical links using both metrics and algorithms, we simulate the energy sharing operation of the cellular network for the connected graphs as obtained in Fig.~\ref{clustering_fig}. Independent realizations of the generated RE $\boldsymbol{\alpha}$ as well as the energy consumption $\mathbf{C}$ of each BS following the distributions specified in Section~\ref{sim_parameters} are used to obtain the energy transactions and the net energy cost of the network. The results are then averaged over $1000$ iterations. Note that to avoid ambiguity between the use of SEA metric for clustering and considering randomness in the day-to-day cost minimization, we have used the perfect knowledge case in these simulations. A summary of the averaged results obtained by sweeping the energy sharing range $r$ is presented in Fig.~\ref{3}.

The energy sharing range directly controls the number of physical links that can be formed in the network. If $r$ is in the range $[0 , 0.5]$ km, then there are no linkages among BSs as all BSs are separated from each other by at least $500$ m. Hence, these can be used as reference points to measure the effectiveness of installing physical energy sharing links among BSs. As the value of $r$ is increased, more energy sharing links can be installed and hence higher utilization of excess RE can be achieved. It can be observed from Fig.~\ref{3a} that increasing the energy sharing range leads to a reduction in the average net energy cost of the operator. This is mainly due to the increasing role of physical energy sharing links that reduces the energy transactions over SG. Hence, the cost gain comes from the reduction in the price paid to SG for the virtual transport of energy. This can be seen from the opposite behaviour in Fig.~\ref{3c} and Fig.~\ref{3e}, i.e., the energy shared over SG decreases while the energy shared over physical links increases as the energy sharing range is increased. Note that the average amount of electricity procured from SG as well as the average amount of electricity sold back to SG remains almost the same as shown by Fig.~\ref{3b} and Fig.~\ref{3d}, respectively. This is because the energy consumption of the network does not change as a result of varying $r$.
Increasing $r$ only encourages the BSs to share more energy over physical links instead of SG, hence, resulting in lower cost.

While comparing the algorithms and metrics, it can be seen that the lowest average energy cost is achieved by the agglomerative-SEA strategy while the highest cost is achieved by the agglomerative-AEA strategy. Although, the divisive-SEA strategy was supposed to perform the best in terms of cost, however, it turns out that it does not lead to the most number of established links. This is because the number of links in the divisive algorithm are limited by the MST approach that prohibits loops and cliques within the graph while such restrictions are not present in the agglomerative approach. Therefore, the agglomerative-SEA strategy ultimately leads to more number of links if the energy sharing range is high enough as it is not constrained by the structure of the graph. The divisive-SEA, on the other hand, performs better in terms of cost reduction as compared to both divisive-AEA and agglomerative-AEA. In order to determine the best combination of metric and algorithm, we need to be aware of the cost of installation of links that is imposed by these strategies. A simple comparison of the total length of links installed using the four strategies (see Fig.~\ref{3g}) reveals that the agglomerative-SEA strategy requires very high length in order to achieve the lowest average energy cost. This makes it less attractive for practical use since the length of installed links is directly related to the cost of installation. The divisive-SEA strategy, on the other hand, performs reasonably well in terms of cost reduction while requiring significantly lower length of installed links. This makes it an attractive strategy to use from a practical perspective.

\vspace{-0.0cm}
\subsection{Comparison of Energy Sharing Strategies} \label{cost_comparison}\vspace{-0.0cm}
In this subsection, we aim to illustrate the difference in the total net energy cost of the network when using different energy sharing strategies and day-to-day optimization techniques. The case of no energy sharing is used as a benchmark for comparison. For the sake of tractability and ease of interpretation, we use a simple example of $K = 3$ BSs labeled as BS 1, BS 2, and BS 3. It is assumed that BS 1 and BS 2 are connected by a physical link of length $2$ km. The third BS is assumed to be isolated and can only share energy via SG. The average amount of RE generated at BS~1 is assumed to be the highest (i.e., 150\% of the maximum consumption) followed by BS~2 (80\% of the maximum consumption) and BS~3 (60\% of the maximum consumption) while the energy consumption of all BSs is assumed to be identical to be able to keep track of the behaviour. The batteries available at BSs sites are assumed to be fully charged at epoch, i.e., $B_{i}(0) = B_{\max}, \forall i \in \{1,\ldots,K\}$.


\begin{figure*}[h!]
\begin{center}
\subfigure[]{\label{cost_benefit}\includegraphics[width=2.6in]{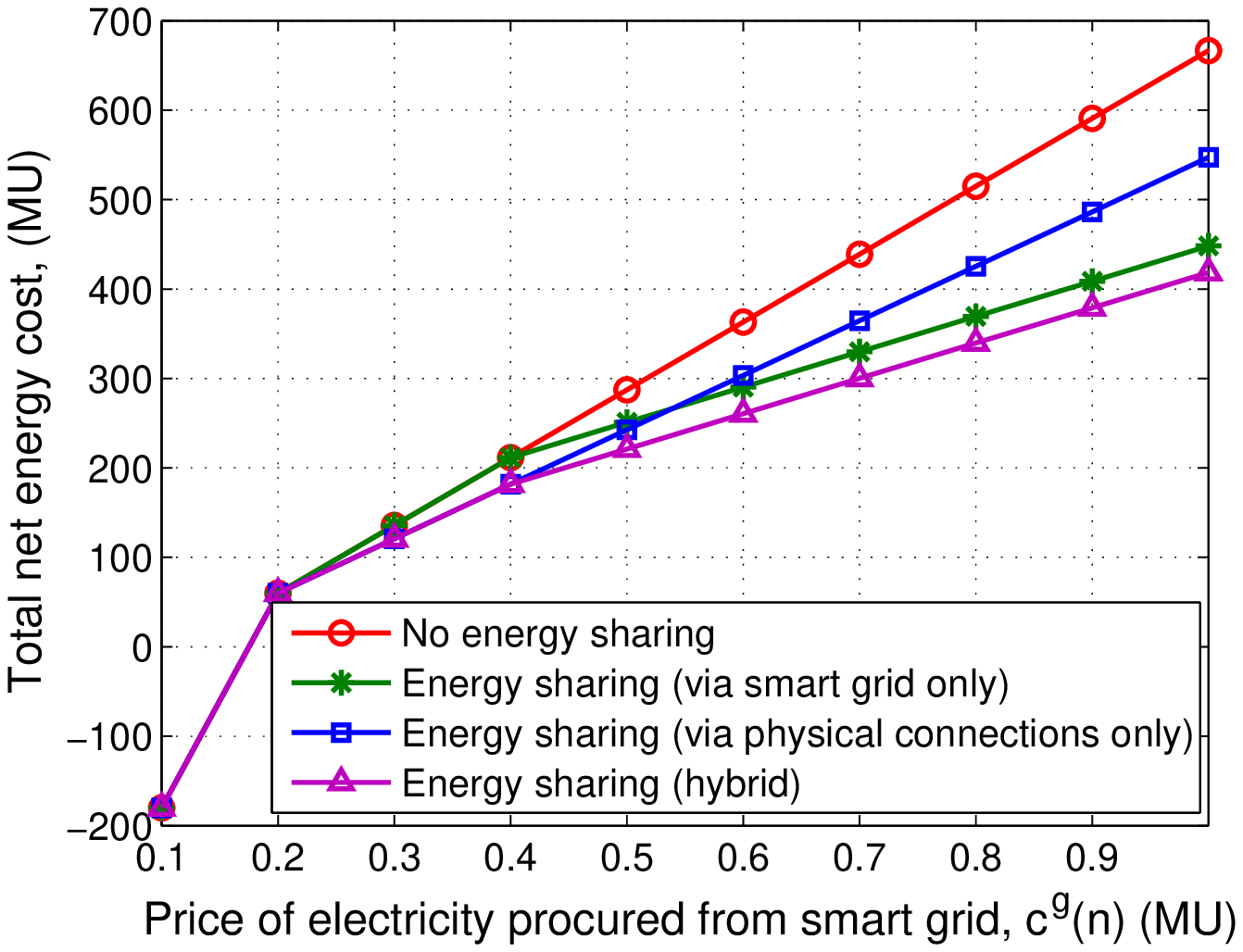}}
\subfigure[]{\label{RE_used}\includegraphics[width=2.6in]{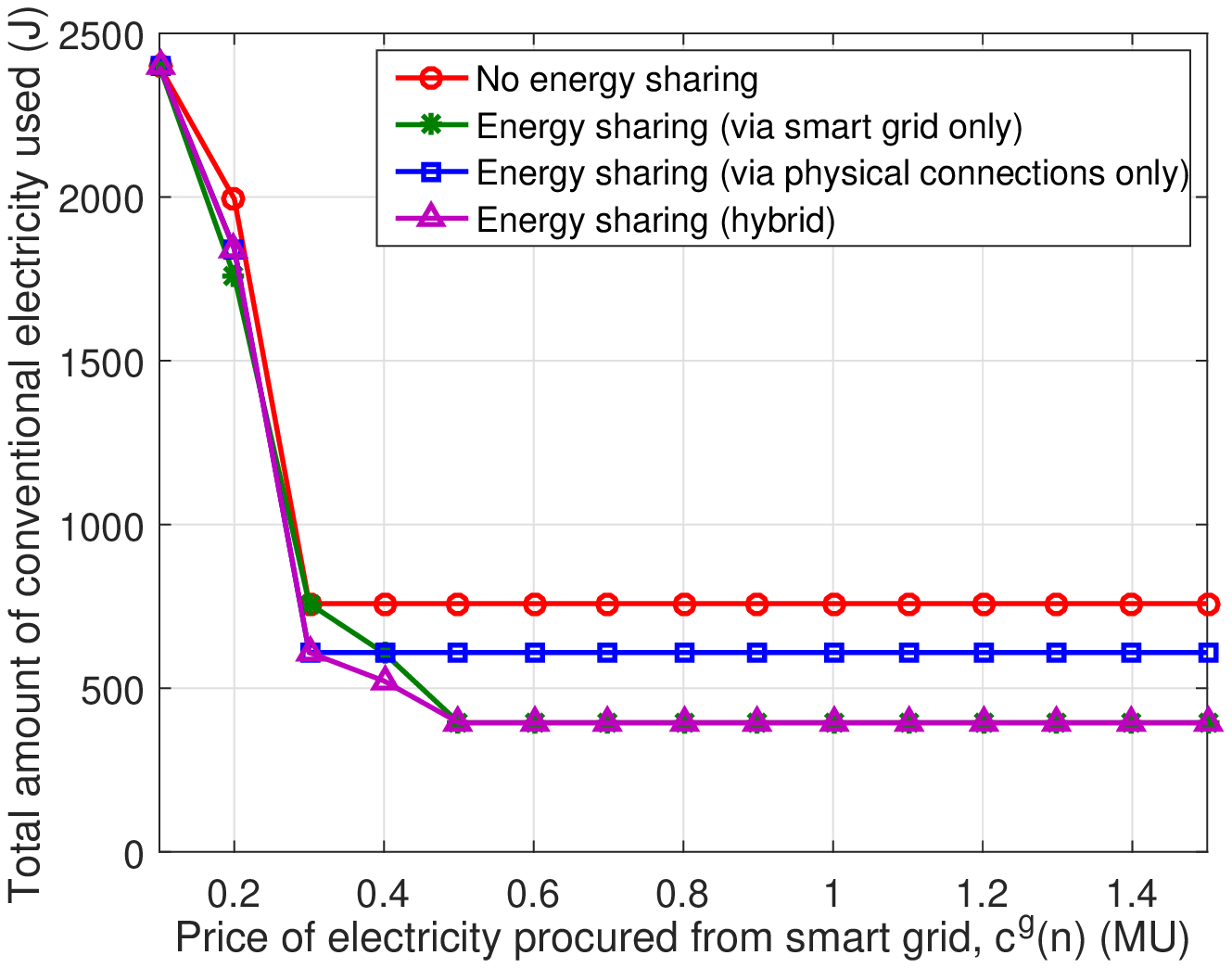}}
\end{center}\vspace{-0.0in}
\caption{Comparison of energy cost and total amount of conventional energy used among different energy sharing strategies.}
\label{Fig_cost}\vspace{-0.1in}
\end{figure*}

The total net energy cost of the network for different energy sharing strategies against increasing electricity price is shown in Fig.~\ref{cost_benefit}. When the selling price of electricity $c^{\text{g}}(n)$ is lower than its buying price $c^{\text{e}}(n) = 0.2$ MU, the BSs are encouraged to sell all the RE generated instead of sharing and buy electricity from SG to fulfill their requirements. As the price of electricity increases beyond $c^{\text{e}}(n)$, the energy sharing starts taking place via physical connections only since it is free as compared to SG that imposes a charge on transporting energy. This explains the behaviour of total net energy cost in Fig.~\ref{cost_benefit} between $c^{\text{g}}(n) = 0.2$ MU and $c^{\text{g}}(n) = 0.4$ MU, i.e., the energy sharing via SG case behaves similar to the case of no energy sharing while the energy sharing via physical links case behaves similar to the hybrid energy sharing case. It is observed that energy sharing has led to a reduction in the total net energy cost of the network. This behaviour ends as the electricity becomes more expensive than buying energy from other BSs using SG, i.e., $c^{\text{b}}(n) = 0.4$ MU. Beyond this point, it can be observed that the lack of energy sharing case, which acts as a benchmark, leads to the highest while the hybrid energy sharing approach leads to the least net energy cost. In this specific scenario of three BSs with a single physical energy sharing link, SG dominates in terms of energy sharing and provides a higher benefit in cost reduction as compared to the physical link. Nevertheless, the difference in the energy sharing via SG case and the hybrid case depends directly on the number of physical energy sharing links in the network. Moreover, in this simple network scenario, the hybrid approach achieves $\approx$ 37\% cost saving as compared to the no energy sharing approach.
\begin{figure*}[h!]
\begin{center}
\subfigure[Deviation in $\boldsymbol{\alpha} = 2\%$]{\label{1pc}\includegraphics[width=2.1in]{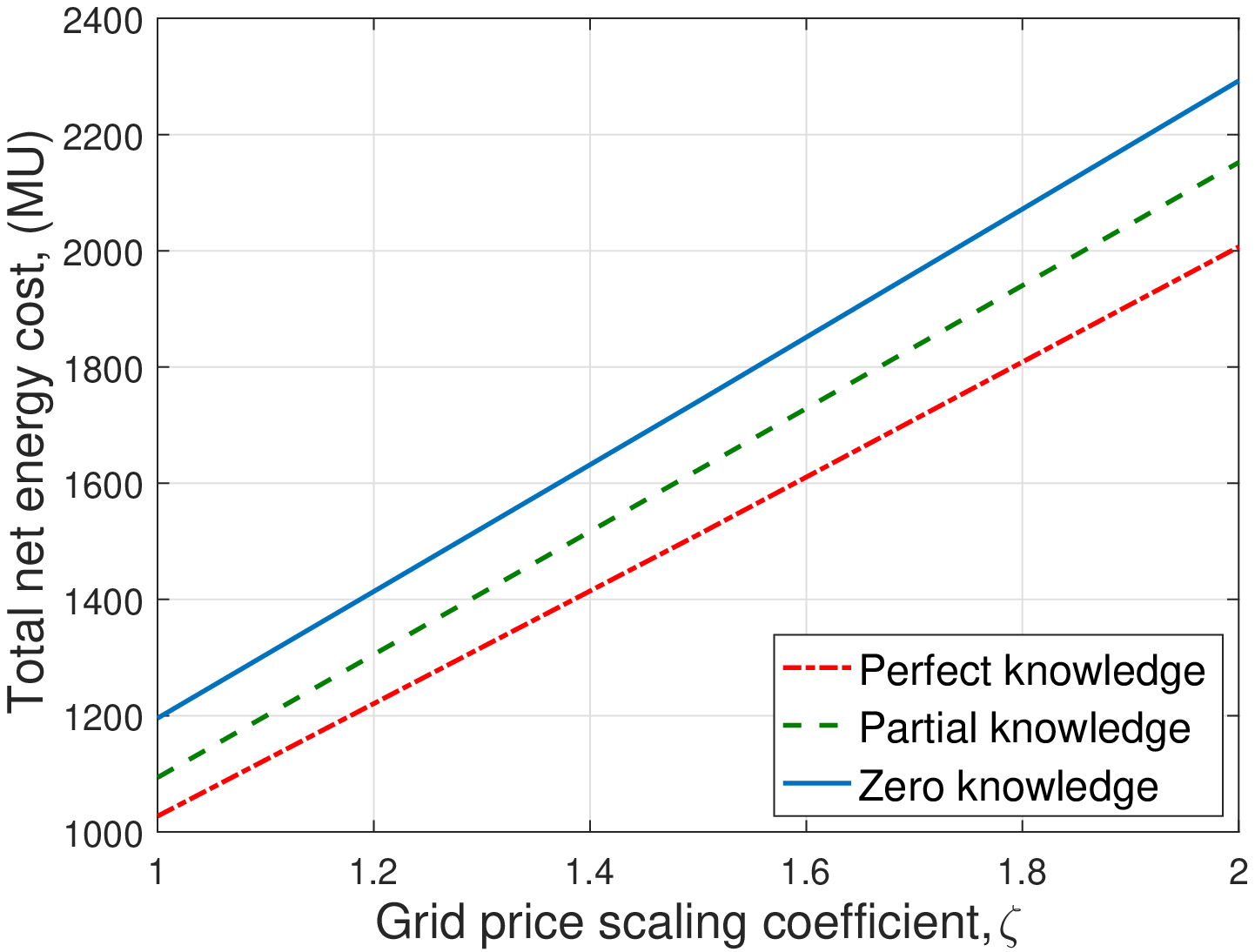}}
\subfigure[Deviation in $\boldsymbol{\alpha} = 5\%$]{\label{5pc}\includegraphics[width=2.1in]{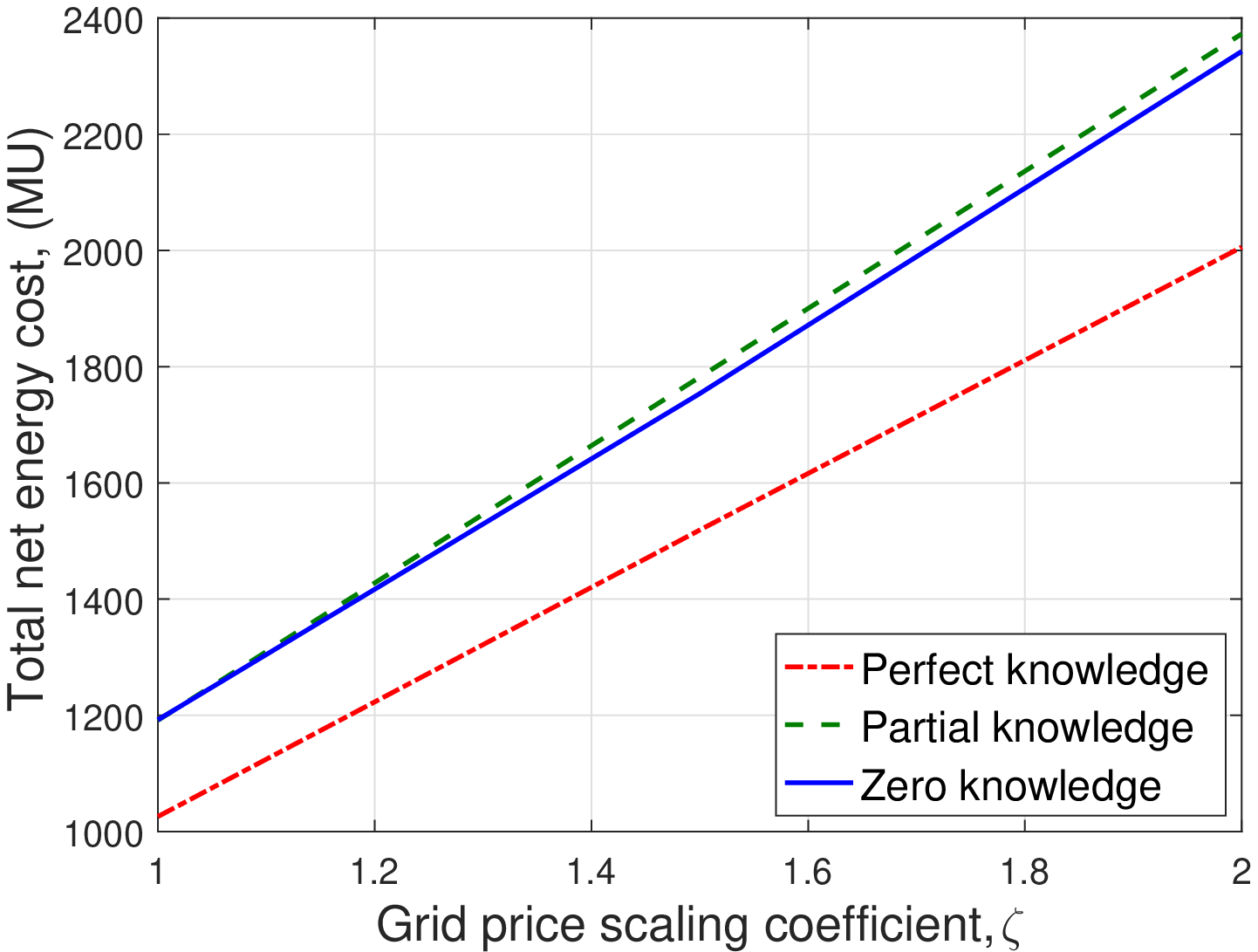}}
\subfigure[Deviation in $\boldsymbol{\alpha} = 10\%$]{\label{10pc}\includegraphics[width=2.1in]{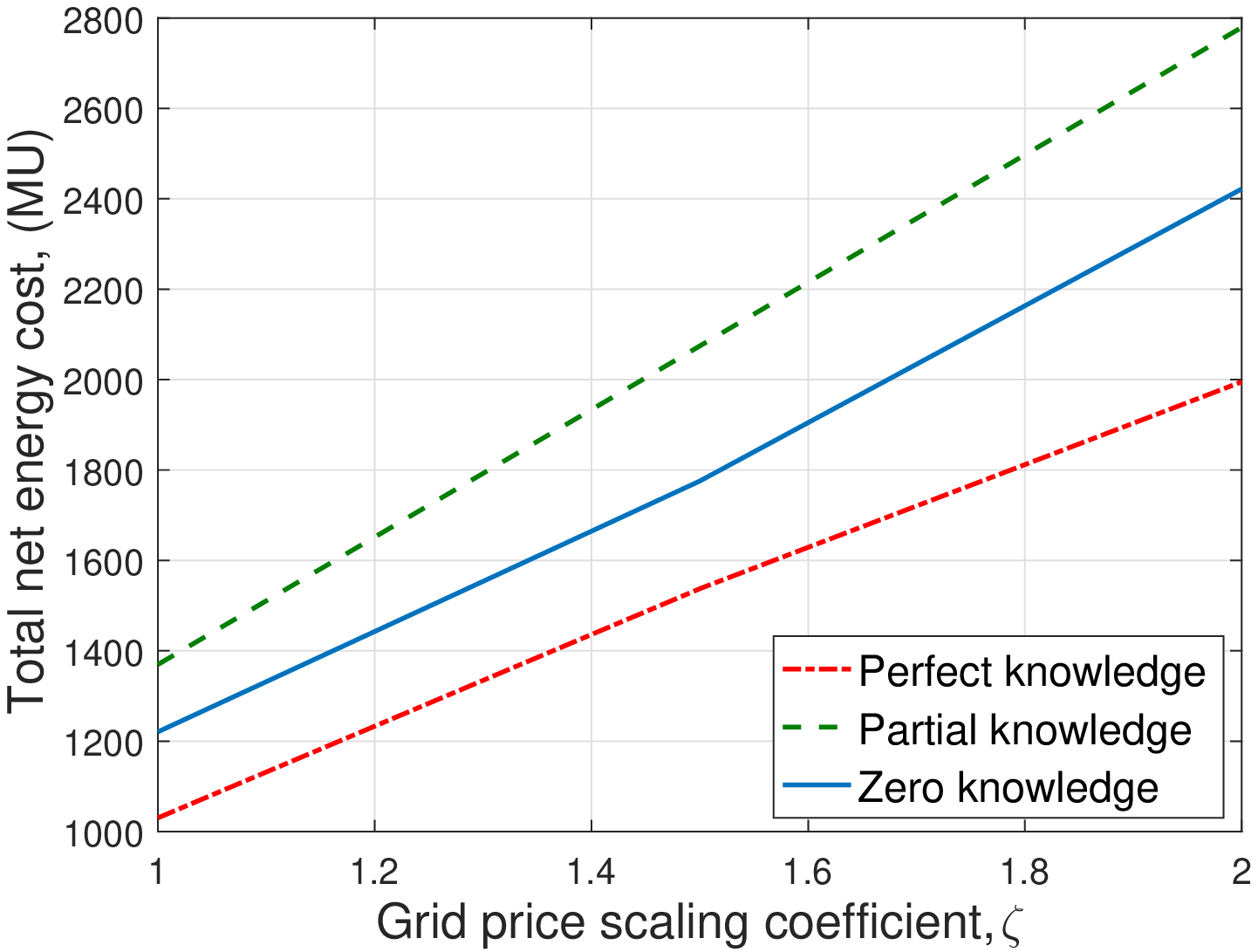}}
\end{center}\vspace{-0.0in}
\caption{Comparison of energy cost for different day-to-day cost optimization strategies.}
\label{Fig_cost_comparison_strategies}\vspace{-0.0in}
\end{figure*}

\textcolor{black}{In Fig.~\ref{RE_used}, we study the amount of conventional electricity bought from SG against the increasing electricity prices for the different strategies. Note that the amount of electricity bought from SG directly reflects the amount of CO$_2$ emissions and hence, the environmental impact of the network. In general, it can be observed that increasing the electricity price of SG $c^g$ reduces the amount of energy procured from the grid for all strategies. This is because of increasing incentive to utilize renewable energy and energy stored in the battery to reduce costs. The proposed hybrid energy sharing scheme performs best in terms of reducing electricity procurement followed by energy sharing via SG and energy sharing via physical connections only. Although the performance of energy sharing via SG and hybrid energy sharing is the same for sufficiently high electricity price, it is important to note that the energy cost incurred by the hybrid scheme is significantly lower. This confirms that physical connections do not increase the use of RE in the system but essentially contribute to achieve additional energy cost reduction.}

\textcolor{black}{Fig.~\ref{Fig_cost_comparison_strategies} shows a comparison of the total net energy cost of the cellular network for the three day-to-day cost minimization strategies, i.e., zero knowledge, perfect knowledge, and partial knowledge. The results are presented for three different deviations from the mean in the actual RE generation. In general, the perfect knowledge case achieves the lowest cost. It can be seen that when the deviations are small, i.e., $2\%$, the partial knowledge case achieves lower cost than that of the zero knowledge case. When the deviations increase to $5\%$, the energy costs of the zero knowledge and perfect knowledge cases are almost the same. However, when the deviations are large, i.e., $10\%$, the partial knowledge case incurs higher energy cost than that of the zero knowledge case since it becomes risk aware and does not completely deplete its available renewable energy.}

\vspace{-0.0cm}
\subsection{Day-to-Day Energy Transactions}\vspace{-0.0cm}
\begin{figure*}[h!]
\addtolength{\subfigcapskip}{-0.0in}
\begin{center}
\subfigure[]{\label{4a}\includegraphics[width=1.9in]{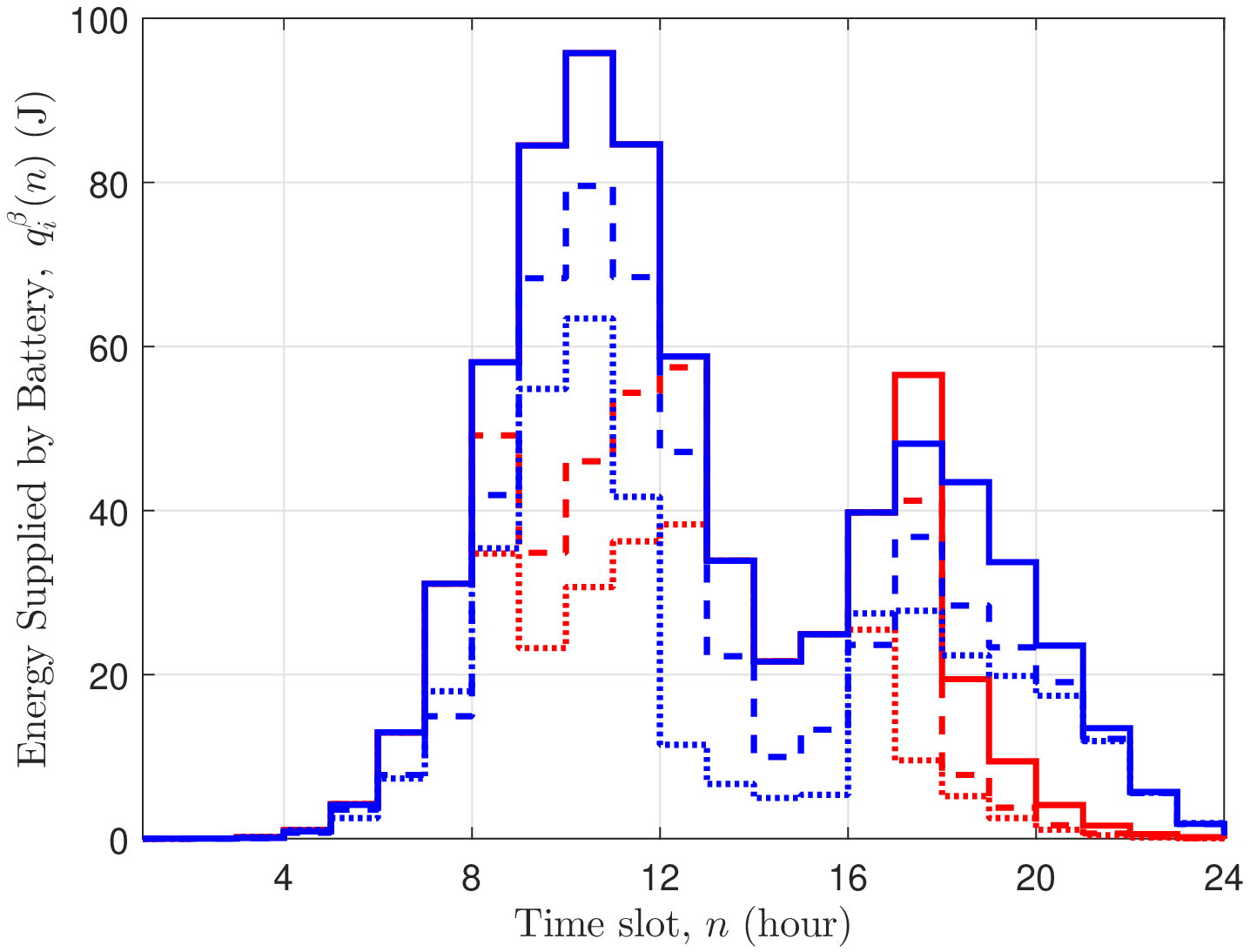}}
\subfigure[]{\label{4b}\includegraphics[width=1.9in]{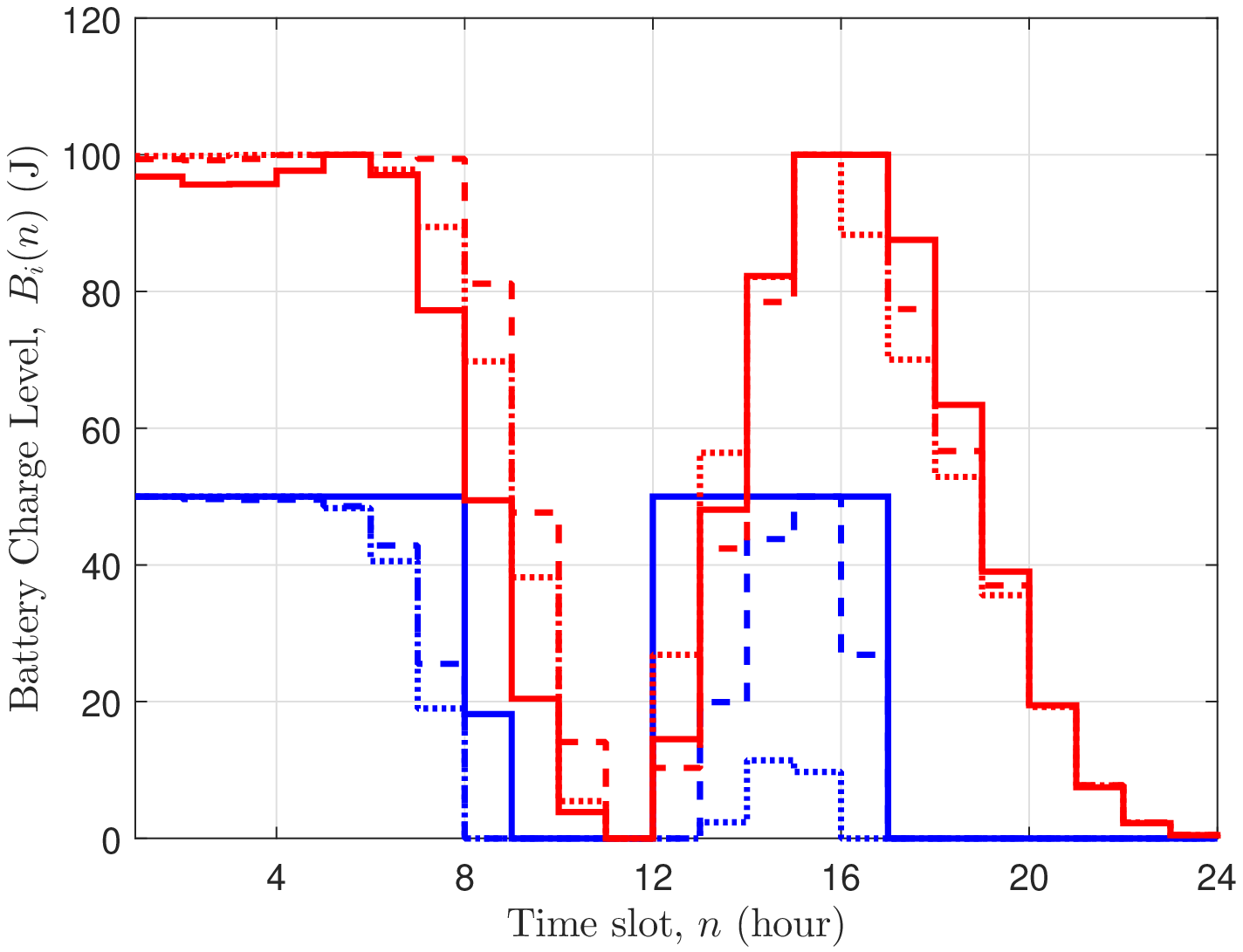}}
\subfigure[]{\label{4c}\includegraphics[width=1.9in]{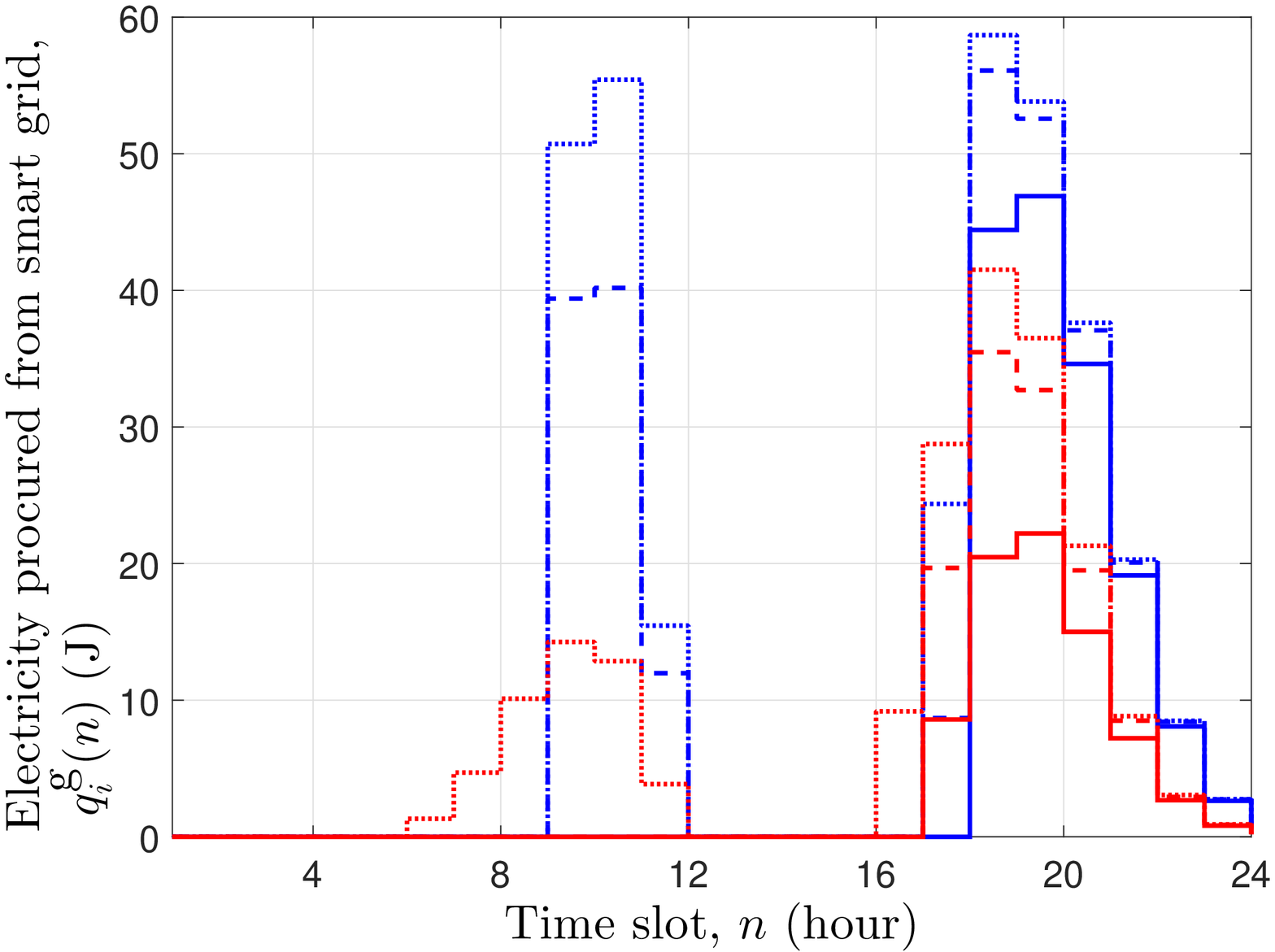}}\\
\subfigure[]{\label{4d}\includegraphics[width=1.9in]{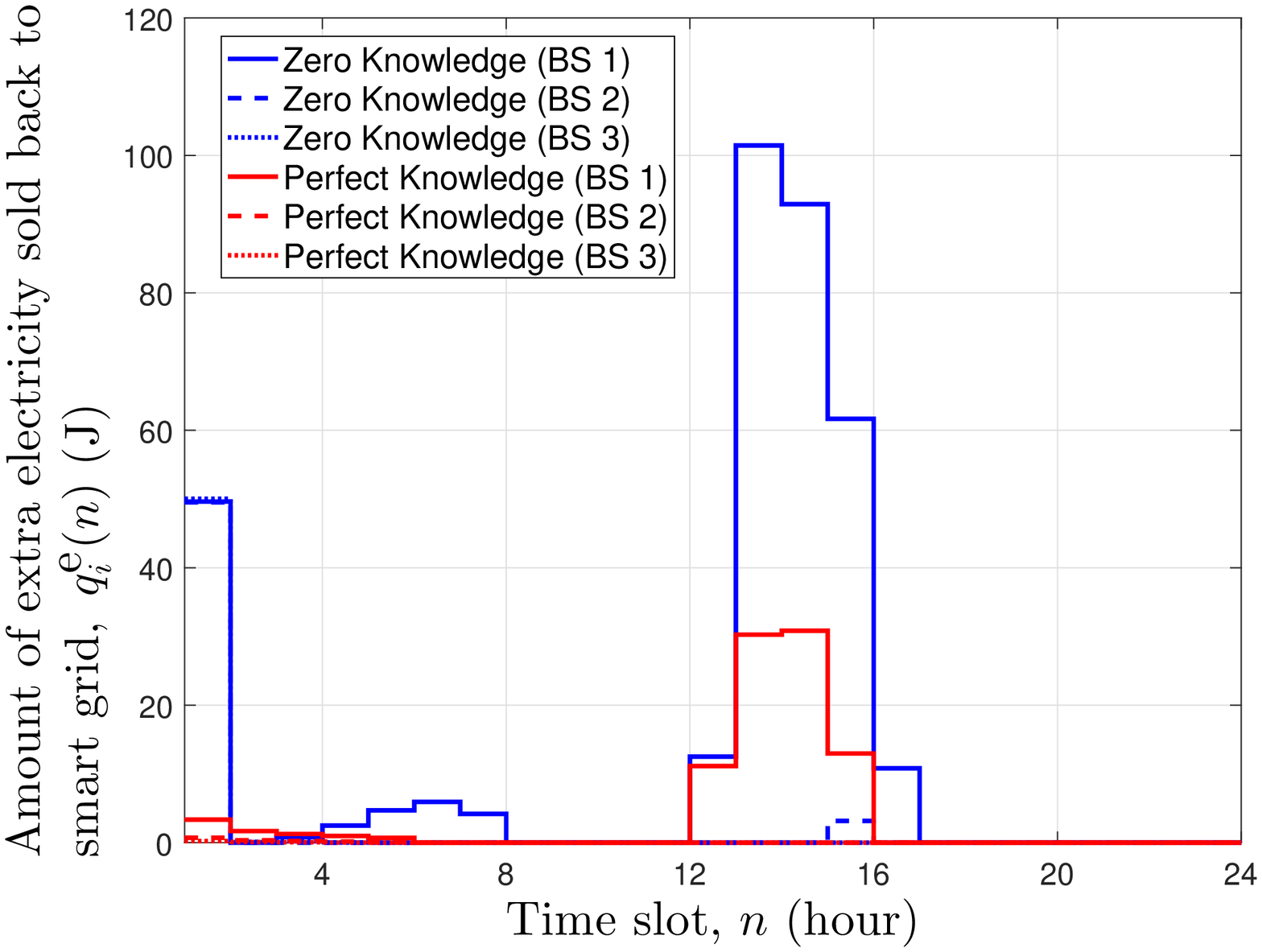}}
\subfigure[]{\label{4e}\includegraphics[width=1.9in]{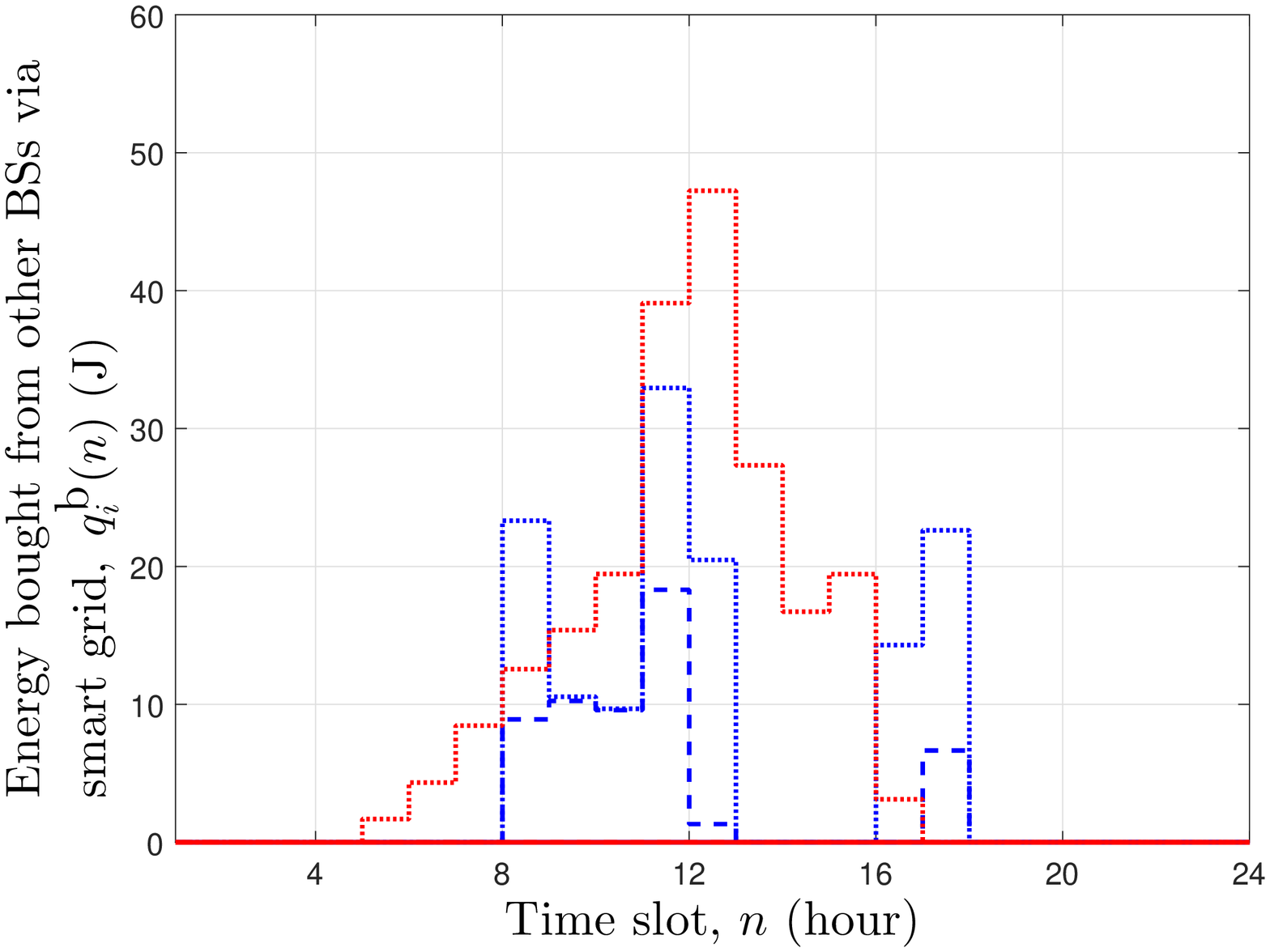}}
\subfigure[]{\label{4f}\includegraphics[width=1.9in]{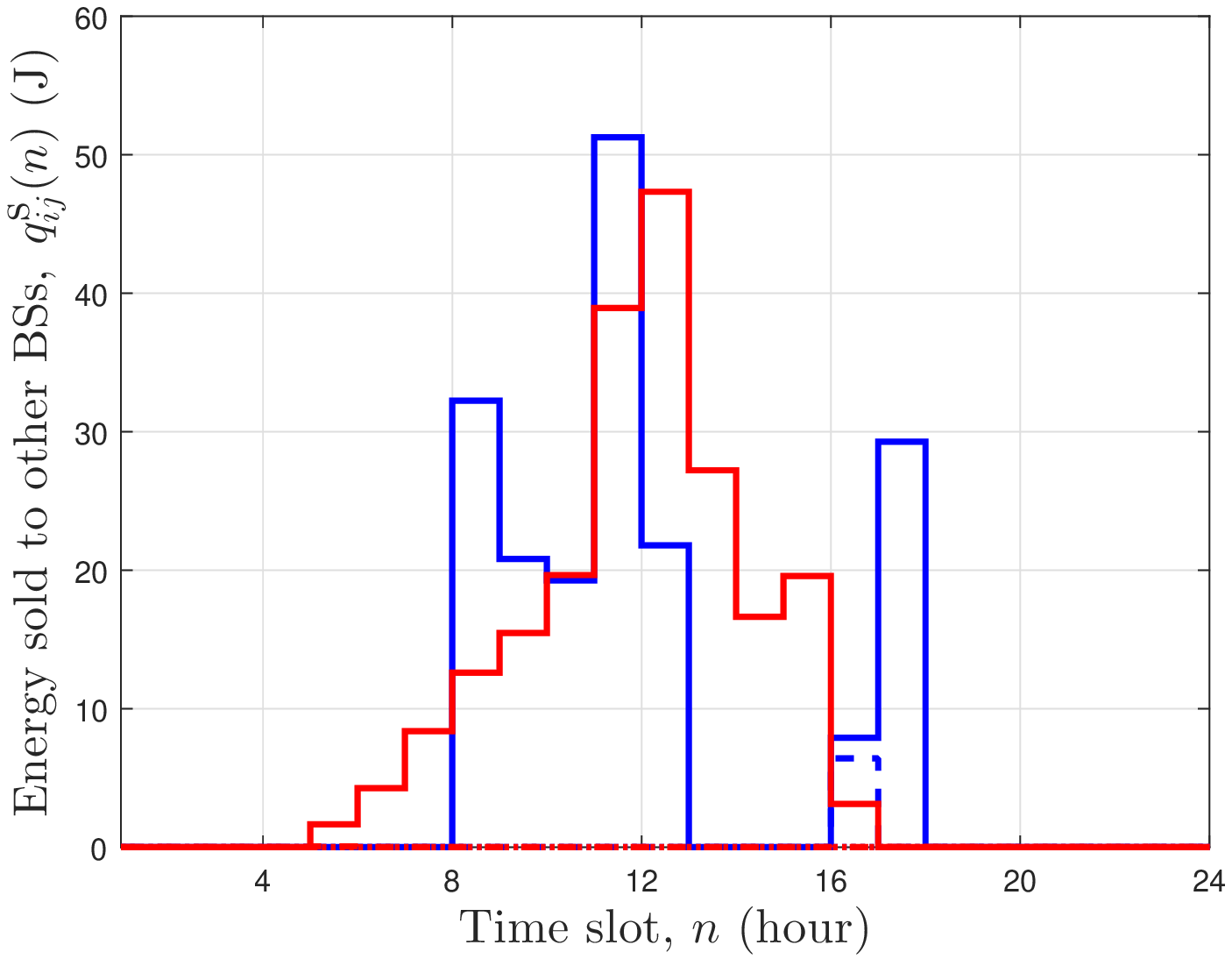}}\\
\subfigure[]{\label{4g}\includegraphics[width=1.9in]{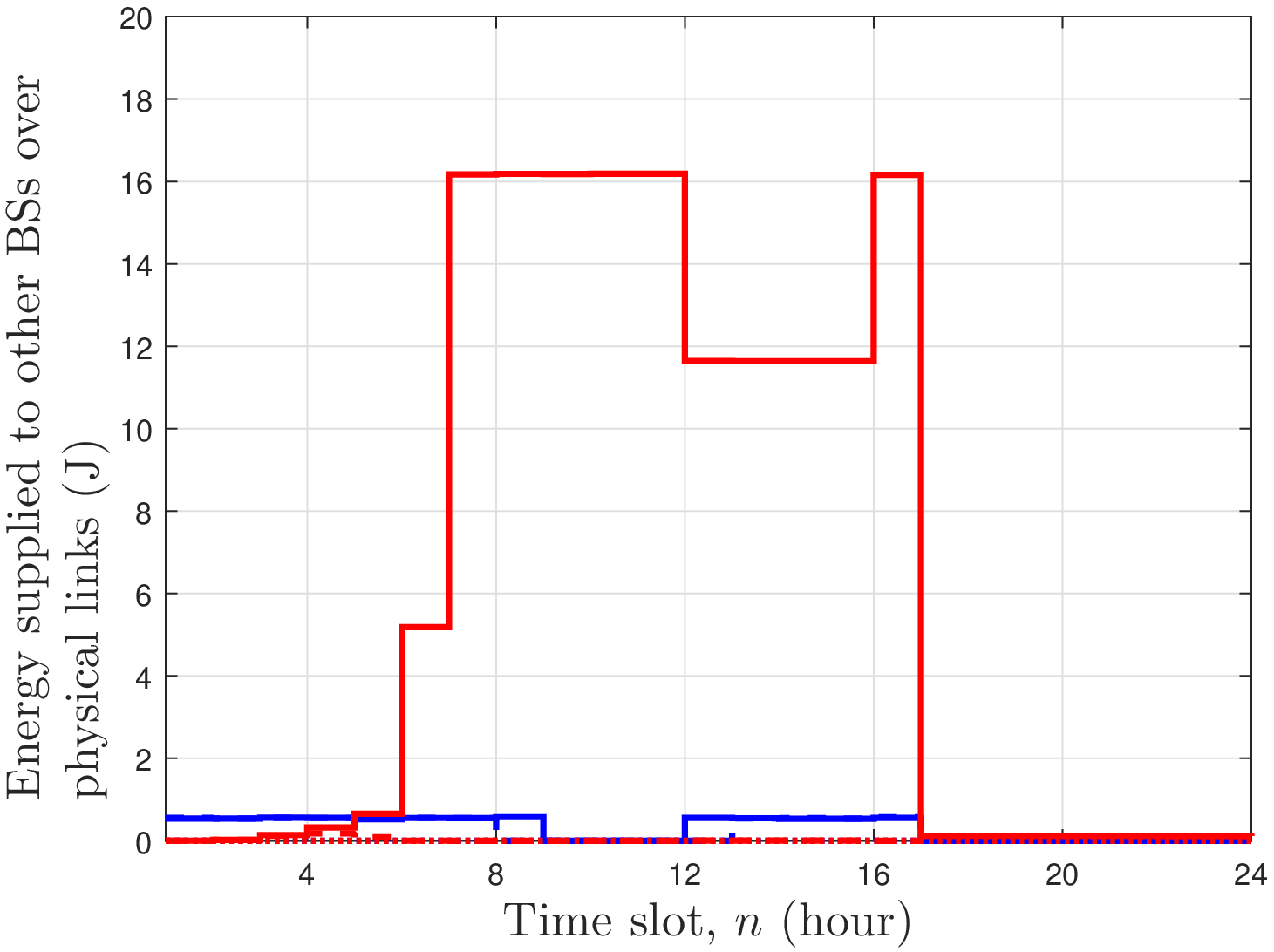}}
\subfigure[]{\label{4h}\includegraphics[width=1.9in]{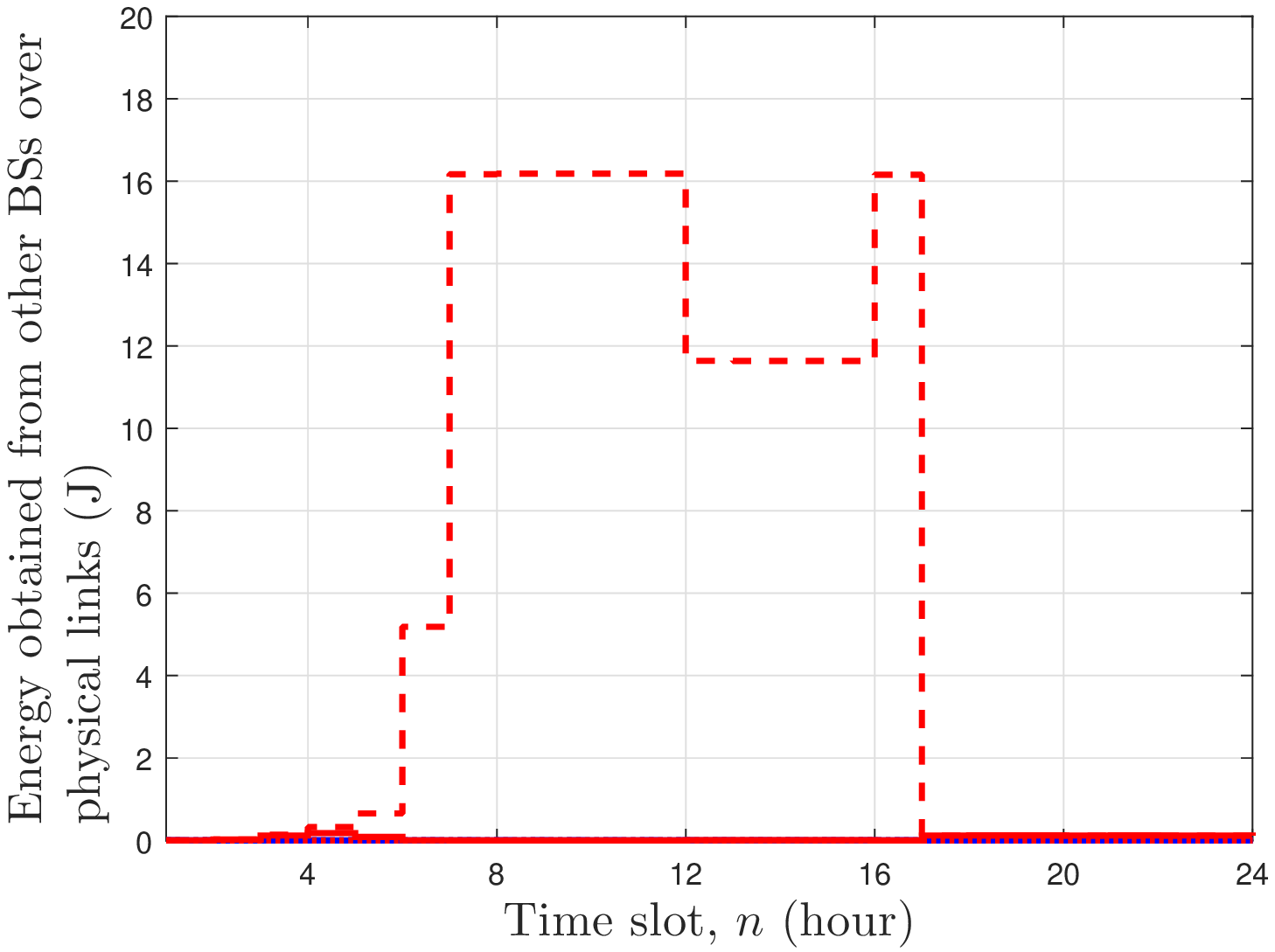}}
\end{center}
\vspace{-0.0in}
\caption{Day-to-day energy transactions of the BSs over time; (a) energy supplied by battery, (b) battery charge level, (c) amount of electricity procured from SG, (d) amount of extra electricity sold to SG, (e) energy bought from other BSs, (f) energy sold to other BSs, (g) energy supplied to other BSs over physical links, and (h) energy obtained from other BSs over physical links. }
\label{Result_vs_time} \vspace{-0.00in}
\end{figure*}

In this section, we use the same simulation setting as used in Section~\ref{cost_comparison} to illustrate the details of the energy transactions and provide a comparison between solutions obtained for the zero knowledge and perfect knowledge cases. A summary of the time varying energy transactions for both cases is provided in Fig.~\ref{Result_vs_time}. Fig.~\ref{4a} shows the energy supplied by the battery to the BS and Fig.~\ref{4b} shows the amount of energy remaining in the battery over time. The amount of energy supplied by the battery follows the consumption profile of the BSs with two peaks at around $10$ am and $6$ pm, respectively. The amplitude of the supplied energy from the battery is in accordance with the available RE. It can be observed from Fig.~\ref{4b} that for the case of zero knowledge of future RE, the battery disposes off any energy in excess of $B_{th} = 50$ Wh from the first time slot. The extra electricity is sold back to SG as shown in the first time slot in Fig.~\ref{4d}. The aggressiveness in using up energy from the battery is due to the fact that in the zero knowledge case, BSs focus only on the current time slot and fail to plan for the future. On the contrary, in the perfect knowledge case, BSs do not deplete the battery from the beginning (as seen by the red lines in Fig.~\ref{4b}). Instead, they saves the energy in the battery for later use during times of high energy consumption and low RE generation.

The electricity supplied by SG and the electricity sold back to SG in the zero knowledge case are significantly higher than that of the perfect knowledge case. This can be observed from Fig.~\ref{4c} and Fig~\ref{4d} with energy being procured during peak consumption times (i.e., 10 am and 6 pm) and extra energy being sold during peak generation times (12~pm). Next, we observe the behaviour of the energy sharing over both physical and virtual links. Due to the network configuration, only BS 3 is involved in buying energy using SG (see Fig.~\ref{4e}) while BS 1 and BS 2 sell the corresponding energy (see Fig.~\ref{4f}) in both zero knowledge and perfect knowledge cases. The energy transactions over physical connections are shown in Fig.~\ref{4g} and Fig.~\ref{4h}. It is clear that the perfect knowledge case allows significant energy exchange over physical links during the peak energy consumption times while the zero knowledge is not involved in sharing energy via physical links. This is due to the better planning and management of energy ahead of time in the perfect knowledge case which is absent in the zero knowledge case.
\begin{figure}[t]
  \centering
  \includegraphics[width=2.6in]{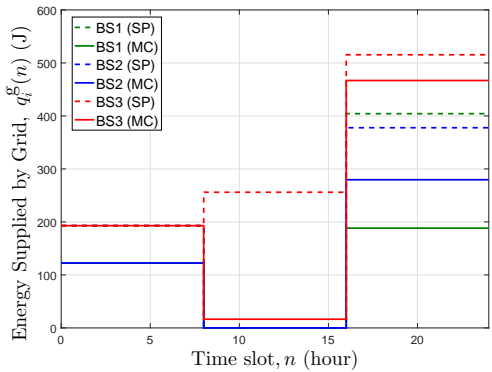}\vspace{-0.0in}
  \caption{Behaviour of energy procurement from the grid under renewable energy uncertainty.}\label{stochastic_fig}\vspace{-0.0in}
\end{figure}

Next, we compare the case of partial knowledge of future RE generation and the case of perfect knowledge averaged over a large number of realizations of the RE $\boldsymbol{\alpha}$. In Fig.~\ref{stochastic_fig}, we plot the average amount of electricity supplied by the grid for all BSs. The dotted lines represent the solution obtained using Monte Carlo (MC) simulations of the perfect knowledge case with 1000 iterations, whereas the solid lines represent the solution obtained from stochastic programming (SP) in the partial knowledge case. For the sake of clarity and ease of understanding, we select N = 3 time slots of 8 hours each. The new set of parameters are selected as follows: the peak traffic times $\mu^{C}_{1}$ = 12 hrs, $\mu^{C}_{2}$ = 20 hrs, the initial battery level $B_{i}(0) = 100$~Wh, $\forall i \in \{1,2,3\}$, and the maximum effective battery capacity $B_{\max}$ = 800~Wh considering the 8 hours time slots. In the first time slot, there is limited RE generation and hence the associated uncertainty is low. Therefore, the SP and MC solutions are very close. Most of the required energy is procured from SG due to lack of RE generation. In the second time slot, the RE generation is high. Therefore, the BSs utilize RE and share the surplus with each other to minimize the procurement of energy from SG. BS 1 and BS 2 have sufficient RE to completely avoid procurement from SG. However, BS 3 requires additional energy from SG to meet its requirements. Notice, however, that the SP solution for BS 3 is much higher than its MC counterpart. This is because there is higher uncertainty in its generation and the SP solution is affected by the worst case situation to avoid the risk. Finally, in the third time slot, the RE generation decreases. Hence, the BSs are encouraged to procure higher amounts of energies from SG. The gap in the SP and MC solutions depicts the risk associated with each decision and depends on the amount of the available RE. 


%
\vspace{-0.0cm}
\section{Conclusions} \label{sec_conclusion}\vspace{-0.0cm}
In this paper, we proposed a hybrid energy sharing framework for cellular networks that are powered by smart grid and have renewable energy generation capabilities. The energy sharing takes place via physical power lines infrastructure as well as the smart grid for virtual energy transportation. Agglomerative and divisive hierarchical clustering algorithms are provided to determine the physical links to be installed based on two different metrics, i.e., average and stochastic energy affinity. After determining the physical connections among BSs, an optimization framework for day-to-day cost optimization is developed for the cases of zero knowledge, perfect knowledge, and partial knowledge about renewable energy generation in the future. The performance of physical connections obtained using the four clustering approaches is compared assuming perfect knowledge of renewable energy generation. The agglomerative algorithm using the stochastic energy affinity metric performs the best in striking a balance between the cost reduction achieved and the initial investment required in installing the links. \textcolor{black}{A comparison is also made between the different day-to-day cost minimization strategies. It is observed that the perfect knowledge case, which is used as a benchmark, performs the best in terms of the energy cost. For the partial knowledge case, the energy cost is higher than that of the perfect knowledge case but lower than that of the zero knowledge case when the uncertainty in the renewable energy generation is low. However, when the uncertainty is sufficiently high, the partial knowledge case becomes risk aware and therefore may incur higher cost than the zero knowledge case.}

\bibliographystyle{ieeetran}
\vspace{-0.0cm}
\bibliography{references}

\begin{IEEEbiography}[{\includegraphics[width=1in,height=1.25in,clip,keepaspectratio]{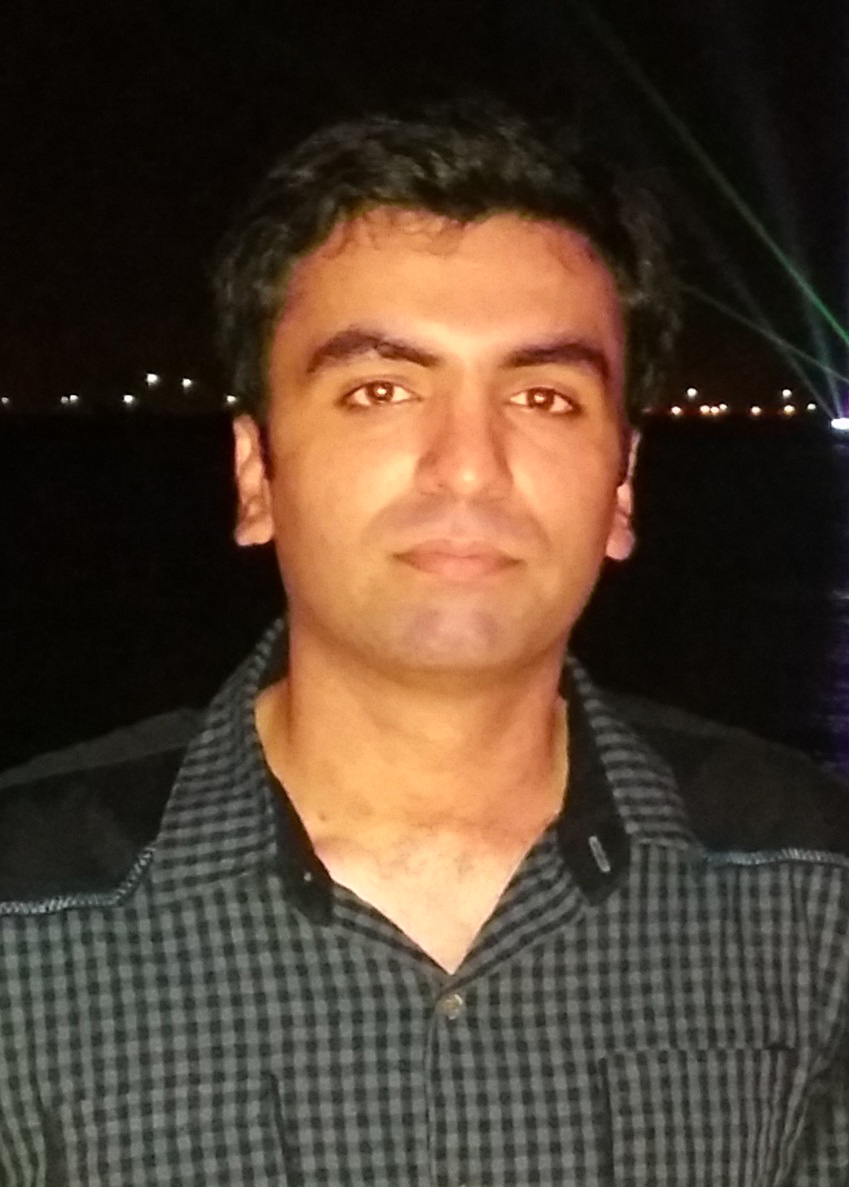}}]{Muhammad Junaid Farooq} (S'15) received the B.S. degree in electrical engineering from the School of Electrical Engineering and Computer Science (SEECS), National University of Sciences and Technology (NUST), Islamabad, Pakistan, the M.S. degree in electrical engineering from the King Abdullah University of Science and Technology (KAUST), Thuwal, Saudi Arabia, in 2013 and 2015, respectively. Then, he was a Research Assistant with the Qatar Mobility Innovations Center (QMIC), Qatar Science and Technology Park (QSTP), Doha, Qatar. Currently, he is a PhD student at the Tandon School of Engineering, New York University (NYU), Brooklyn, New York. His research interests include modeling, analysis and optimization of wireless communication systems, stochastic geometry, and green communications. He was the recipient of the President's Gold Medal for the best academic performance from the National University of Sciences and Technology (NUST).
\end{IEEEbiography}

\begin{IEEEbiography}[{\includegraphics[width=1in,height=1.25in,clip,keepaspectratio]{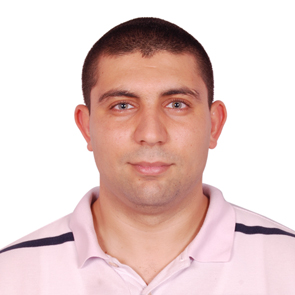}}]{Hakim Ghazzai} (S'12, M'15)
was born in Tunisia. He is currently working as a research scientist at Qatar Mobility Innovations Center (QMIC), Doha, Qatar. He received his Ph.D degree in Electrical Engineering from King Abdullah University of Science and Technology (KAUST), Saudi Arabia in 2015. He received his Diplome d'Ingenieur in telecommunication engineering and Master of Science degree from the Ecole Superieure des Communications de Tunis (SUP'COM), Tunisia in 2010 and 2011, respectively. His general research interests include mobile and wireless networks, green communications, internet of things, UAV-based communications, and optimization.
\end{IEEEbiography}

\begin{IEEEbiography}[{\includegraphics[width=1in,height=1.25in,clip,keepaspectratio]{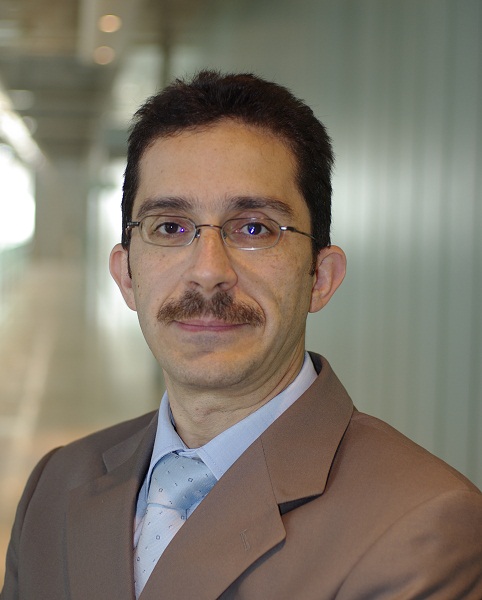}}]{Abdullah Kadri} (SM'16)
received the M.E.Sc. and Ph.D. degrees in electrical engineering from the University of Western Ontario (UWO), London, ON, Canada, in 2005 and 2009, respectively. Between 2009 and 2012, he worked as a Research Scientist at Qatar Mobility Innovations Center (QMIC), Qatar University. In 2013, he became a Senior R$\&$D Expert and the Technology Lead at QMIC focusing on R$\&$D activities related to intelligent sensing and monitoring using mobility sensing. His research interests include wireless communications, wireless sensor networks for harsh environment applications, indoor localization, internet-of-things, and smart sensing. He is the recipient of the Best Paper Award at the WCNC Conference in 2014.
\end{IEEEbiography}

\begin{IEEEbiography}[{\includegraphics[width=1in,height=1.25in,clip,keepaspectratio]{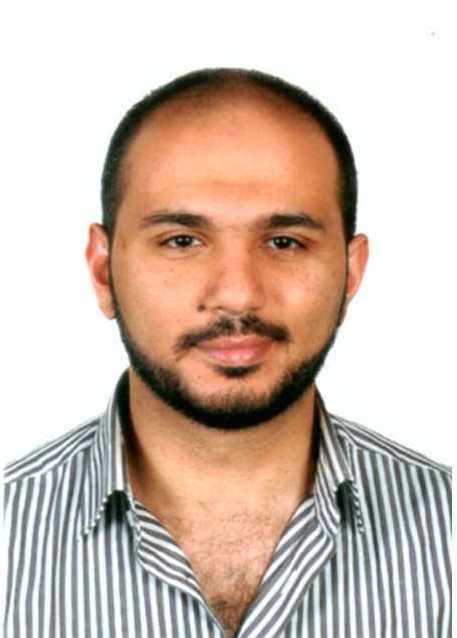}}]{Hesham ElSawy} (S'10, M'14) received the B.Sc. degree in Electrical Engineering from Assiut University, Assiut, Egypt, in 2006, the M.Sc. degree in Electrical Engineering from Arab Academy for Science and Technology, Cairo, Egypt, in 2009, and the Ph.D. degree in Electrical Engineering from the University of Manitoba, Winnipeg, MB Canada, in 2014. Currently, he is a postdoctoral fellow with the Computer, Electrical, and Mathematical Sciences and Engineering Division, King Abdullah University of Science and Technology (KAUST), Saudi Arabia, and an adjunct faculty at the school of Computer Science \& Engineering, York University, Canada. During the period of 2006-2010, he worked at the National Telecommunication Institute, Egypt, where he conducted professional training both at the national and international levels, as well as research on network planning. From 2010 to 2014, he worked with TRTech, Winnipeg, MB, Canada, as a Student Researcher. For his academic excellence, he has received several academic awards, including the NSERC Industrial Postgraduate Scholarship during the period of 2010-2013, and the TRTech Graduate Students Fellowship in the period of 2010-2014. He also received the best paper award in the ICC 2015 workshop on small cells and 5G networks. He is recognized as an exemplary reviewer by the {\em IEEE Transactions of communication}. His research interests include statistical modeling of wireless networks, stochastic geometry, and queueing analysis for wireless communication networks.
\end{IEEEbiography}

\begin{IEEEbiography}[{\includegraphics[width=1in,height=1.25in,clip,keepaspectratio]{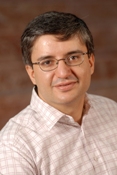}}]{Mohamed-Slim Alouini}
(S'94, M'98, SM'03, F'09) was born in Tunis, Tunisia. He received the Ph.D. degree in Electrical Engineering from the California Institute of Technology (Caltech), Pasadena, CA, USA, in 1998. He served as a faculty member in the University of Minnesota, Minneapolis, MN, USA, then in the Texas A \& M University at Qatar, Education City, Doha, Qatar before joining King Abdullah University of Science and Technology (KAUST), Thuwal, Makkah Province, Saudi Arabia as a Professor of Electrical Engineering in 2009. His current research interests include the modeling, design, and performance analysis of wireless communication systems.
\end{IEEEbiography}

\end{document}